\documentstyle[manuscript,aps,epsfig]{revtex}

\begin{document}

\title{Hamiltonian Theory of the Composite Fermion Wigner Crystal}

\author{R. Narevich, Ganpathy Murthy and H. A. Fertig}

\address{Department of Physics and Astronomy, University of Kentucky, 
Lexington, KY 40506-0055}

\date{\today}

\maketitle 
\begin{abstract}

Experimental results indicating the existence of the high magnetic
field Wigner Crystal have been available for a number of years. While
variational wavefunctions have demonstrated the instability of the
Laughlin liquid to a Wigner Crystal at sufficiently small filling,
calculations of the excitation gaps have been hampered by the strong
correlations. Recently a new Hamiltonian formulation of the fractional
quantum Hall problem has been developed. In this work we extend the
Hamiltonian approach to include states of nonuniform density, and use
it to compute the excitation gaps of the Wigner Crystal states. We
find that the Wigner Crystal states near $\nu=1/5$ are quantitatively
well described as crystals of Composite Fermions with four vortices
attached. Predictions for gaps and the shear modulus of the crystal
are presented, and found to be in reasonable agreement with
experiments.
\end{abstract}

\section{INTRODUCTION AND PREVIEW}

The Fractional Quantum Hall {(}FQH{)} regime presents us with the 
quintessential
problem of strong correlations. In a strong magnetic field $B$, the
kinetic energy of the two-dimensional electron gas (2DEG) is quantized
into Landau levels with energy $(n+{1\over2})\omega_c$, where
$\omega_c=eB/m$ is the cyclotron frequency. Each of these Landau
levels (LLs) has a huge degeneracy equal to the number of quanta of
flux penetrating the 2DEG. When the lowest Landau level (LLL) is
partially full, it is seen that the kinetic energy is degenerate, and
all the dynamics has to come from interactions. At certain special
filling factors (recall that filling factor is \( \nu =2\pi n/eB \))
 the system reorganizes itself into new strongly
correlated ground states\cite{fqhe-ex} with fractionally charged
excitations\cite{laugh}.

The past decade has seen the development and acceptance of the
Composite Fermion {(}CF{)} concept as basic to the understanding of a
variety of these electronic states\cite{jain-book}.  
The CF is pictured as an electron
bound to an even number $l$ of quanta of statistical flux, which are
opposed to the external field. At the mean field level, each CF sees
both the external field and the statistical field due to the other
particles, and therefore moves in an effective field $B^*=B-2\pi l n$,
where $n$ is the density of electrons. The principal fractions
$\nu=p/2p+1$ are seen to be exactly those fillings when the number of
particles is exactly enough to fill an {\it integer} number of LLs of
the effective field. 

Thinking in terms of CFs greatly simplifies the description of
different incompressible and compressible FQH states. CFs are believed
to be the true quasiparticles in much the same way as Landau
quasiparticles are for the normal Fermi liquid.  Following Laughlin's
seminal insight\cite{laugh}, the original CF theory was based on
generating electronic wave functions\cite{jain}.  Contemporaneously,
field-theoretic approaches\cite{field-bos} were also developed to
better understand the FQHE, and to compute response functions. Most of
the field-theoretic approaches are based on the Chern-Simons (CS)
transformation, a method of attaching flux to particles in two
dimensions. Attaching an odd number of flux quanta to electrons
transforms them into bosons and leads to the bosonic CS
approach\cite{field-bos}, while adding an even number leads to the
fermionic CS theory\cite{field-theory}.  In the mean field
approximation the fermionic CS theory recovers the picture of CFs in
an effective magnetic field.  Recently, based on the fermionic CS
theory, a Hamiltonian approach was developed\cite{shankar-murthy} 
to describe liquid states in the FQH
regime.  In this approach, the CF representation is reached from the
bare electronic coordinates by a series of canonical
transformations. The end product is the electron density operator
reexpressed in the CF coordinates suitable for further calculations
and/or approximations. Physical quantities calculated in this
approach seem to be in reasonable agreement with numerical results and
experiments\cite{sm-work}.

The subject of this paper, however, is not the liquid FQH states, but
the insulating states that have been detected experimentally at very
low filling fractions\cite{exp_rev}.  A natural candidate to exhibit
such insulating behavior is the electronic Wigner Crystal
{(}WC{)}. The simplest description of this state is the Hartree-Fock
{(}HF{)} wave function\cite{maki}
\begin{equation}
\label{HF-trial-wf}
\Psi _{HF}(\{{\mathbf{r}}_{i}\})={\mathcal{A}}\prod _{i}\phi _{{\mathbf{R}}_{i}}({\mathbf{r}}_{i}).
\end{equation}
where \( {\mathcal{A}} \) is the antisymmetrization operator, and \(
\phi _{{\mathbf{R}}_{i}} \) is a single-particle wave function that is
localized at \( {\mathbf{R}}_{i} \) (lattice site) and belongs to the
LLL. It is given by
\begin{equation}
\label{single-particle}
\phi _{{\mathbf{R}}_{i}}({\mathbf{r}})=e^{-\left| {\mathbf{r}}-{\mathbf{R}}_{i}\right| ^{2}/4l_{0}^{2}-i{\mathbf{r}}\times {\mathbf{R}}_{i}\cdot \hat{z}/2l_{0}^{2}},
\end{equation}
 where \( l_{0}=(eB)^{-1/2} \) is the magnetic length. The wave
function (\ref{HF-trial-wf}) has been improved by adding a Jastrow
correlation factor\cite{lam}, and the energy of the resulting state
has been shown to become lower than that of the liquid state at about
the experimentally right filling fraction (\( \nu \approx \frac{1}{7}
\)) \cite{maki,levesque,lam}. Thus, a very strong magnetic field
favors crystalline order by localizing the electrons.

However, not all the experimental evidence supports the simple
electronic WC picture. In particular, transport
experiments\cite{jiang1,jiang2} suggest that the activation gap in
this system is two orders of magnitude smaller than the theoretical
estimate as calculated using the Hartree-Fock
approximation\cite{fukuyama,yoshioka}.  Moreover, close to the
Laughlin fractions \( \nu =\frac{1}{2p+1} \), a dip in the
longitudinal resistivity \( \rho _{xx} \) is observed\cite{goldman1},
resembling the behavior of the correlated liquid state. The
measurements of the Hall resistivity \( \rho _{xy} \) are surprising
as well\cite{goldman1,goldys,goldman2}. The electronic WC is known to
have a vanishing Hall conductance $\sigma_{xy}=0$, which implies a
vanishing Hall resistance $\rho_{xy}=0$.  On the contrary, experiments
see Hall insulating behavior, that is, $\rho_{xy}\approx {h\over \nu
e^2}$. These problems led Yi and Fertig\cite{yi-fertig} to consider
crystalline states with Laughlin-Jastrow correlations. The idea is to
construct a trial wave function with correlation factors that keep
electrons apart. Each electron is combined with \( l \) vortices to
obtain the trial wave function
\begin{equation}
\label{trial wf}
\Psi (\{{\mathbf{r}}_{i}\})={\mathcal{A}}\prod _{i\neq j}(z_{i}-z_{j})^{l}\prod _{i}\phi _{{\mathbf{R}}_{i}}({\mathbf{r}}_{i}).
\end{equation}
The Coulomb energy for this wave function was then computed using Monte
Carlo simulation. Yi and Fertig showed that the ground state energy of
the correlated WC is lower than that of the usual WC at experimentally
relevant filling fractions\cite{yi-fertig}.  Moreover by introducing
Laughlin-Jastrow correlations between the interstitials and the
lattice electrons the experimentally observed \( \rho _{xy} \) (Hall
insulating behavior)\cite{zheng1,zheng2} can be
explained. Unfortunately, the method becomes too computationally
demanding to allow one to calculate other quantities of interest, such
as the excitation spectrum. 

Since the Laughlin-Jastrow correlations are precisely the ones that
convert electrons into CFs\cite{jain} we are led to consider a crystal
of Composite Fermions. The main advantage of the Hamiltonian approach
is that one can easily compute the excitation gap along with the
ground-state energy.

Our main result is that the CF theory with \( l=4 \) zeros attached to
each particle gives the best description of the experimental
phenomenology. Figure 1 shows the results of our calculations of the
excitation gap \( E_{g} \) as a function of the filling factor around
\( \nu =1/5 \) in the triangular lattices of CFs.  Our theory
reproduces the dependence of the excitation energies on the filling
factors as measured in\cite{jiang2} reasonably well when \( \nu <1/5
\). We also find that the shear modulus of the CF lattices goes down
as the filling factor \( \nu =1/5 \) is approached from below.  This
behavior is consistent with the experimentally observed increase of
the threshold voltage for filling factors \( \nu \rightarrow 1/5 \)
\cite{engel} (if one interprets the results in terms of ``weak
pinning''\cite{fukuyama-lee}). 

For $\nu>1/5$ we will show in section V that the energy landscape
becomes very flat, with many different lattice structures becoming
nearly degenerate in energy. Not coincidentally, the convergence of
our Hartree-Fock procedure is very poor in this region, and we are
unable to identify the proper ground state in the clean limit. We have
therefore presented two values for the gaps in this region, the upper
one being for the triangular lattice, and the lower one being for a
more oblique lattice. Neither gap follows the experimentally observed
non-monotonic dependence \( E_{g}(\nu ) \).  (Note, however, the
different slope below and above \( \nu =1/5 \) in Fig. 1). We believe
the main reason for this is the following: Since there are many local
minima with different lattice structures which are very close in
energy, disorder may play an important role in real samples. The
experimental gaps may also be dominated in this region by disorder
effects. Apart from this one region of discrepancy, our numbers for
the gaps are in reasonable agreement with experiments.

The outline of the paper is as follows. In Section II we introduce the
Hamiltonian formalism and show how the wave function (\ref{trial wf})
emerges naturally in this approach. In Section III we derive the
expression of the electronic density operator in the CF representation
and also calculate the Hall conductance for the CF lattice in the clean
limit. In Section IV we formulate the HF theory using the electronic
density derived in Section III. In section V we present the results
and discuss their physical import. Some details of the
calculations are relegated to the three appendices.

\section{SETTING UP THE HAMILTONIAN FORMALISM}

In this section we follow closely the exposition and notation of
Murthy and Shankar\cite{M-S jain book}. Since most of the details are
similar, we will mainly highlight the differences. We will follow
Lopez and Fradkin\cite{field-theory} in assuming that a good starting
point for obtaining a perturbative solution of the \( N \) interacting
electron problem moving in a uniform magnetic field \(
{\mathbf{B}}=-\hat{z}\, B \) is the non-interacting CS particle
Hamiltonian (\( \hbar =c=1 \), \( m \) is the mass of the electron, \(
e \) its charge)
\begin{equation}
\label{CS Hamiltonian}
H_{CS}=\sum _{i}^{N}\frac{1}{2m}(-i\nabla _{i}+e{\mathbf{A}}^{*}({\mathbf{r}}_{i})+{\mathbf{a}}_{CS}({\mathbf{r}}_{i}))^{2}.
\end{equation}
 The CS particle is obtained from the electron by attaching \( l \) flux quanta
to it. This flux attachment is the origin of the Chern-Simons gauge field \( {\mathbf{a}}_{CS}({\mathbf{r}}_{i}) \),
which is defined by 
\begin{equation}
\label{gauge field}
\frac{\nabla \times {\mathbf{a}}_{CS}({\mathbf{r}}_{i})}{2\pi l}=\sum _{i}^{N}\delta ^{2}({\mathbf{r}}-{\mathbf{r}}_{i})-n({\mathbf{r}}),
\end{equation}
 where \( n({\mathbf{r}}) \) is the expectation value of the electron density
in the ground state (CS particle and electron densities are the same). Unlike
the incompressible FQH state (liquid state) for which the electrons have a uniform
density, we are concerned with the case where the density depends on the
position. The vector potential \( {\mathbf{A}}^{*}({\mathbf{r}}_{i}) \) corresponds
to the difference of the external magnetic field and the average field created
by the attached flux tubes, or \( B^{*}({\mathbf{r}})=B-2\pi ln({\mathbf{r}})/e \).

Our main assumption throughout this work is that for the appropriate
field strength and electron density the electrons organize
themselves into a density wave state such that the ground state
expectation value of the density operator is a periodic function
\begin{equation}
\label{average density}
n({\mathbf{r}})=n+\sum _{{\mathbf{G}}\neq 0}\delta n({\mathbf{G}})e^{i{\mathbf{G}}\cdot {\mathbf{r}}},
\end{equation}
 where \( {\mathbf{G}} \) are the reciprocal lattice vectors. We also
assume that the uniform component of the average density is much
larger than any of the finite-\( {\mathbf{G}} \) modulations, i.e. \(
n\gg \delta n \). Effectively \( \delta n/n \) is a small parameter in
our theory. The effects of disorder are ignored in what follows.

The complicated part of  the CS Hamiltonian in (\ref{CS Hamiltonian}) is
the gauge field \textbf{\( {\mathbf{a}}_{CS}({\mathbf{r}}_{i}) \)}. To get rid
of it, one enlarges the Hilbert space\cite{shankar-murthy} by introducing a
canonical pair of fields \( a({\mathbf{q}}),P({\mathbf{q}}) \) for every \( {\mathbf{q}} \)
\begin{equation}
\label{a,P commutator}
[a({\mathbf{q}}),P({\mathbf{q}}^{\prime })]=i\, (2\pi )^{2}\delta ^{2}({\mathbf{q}}+{\mathbf{q}}^{\prime })
\end{equation}
where \( {\mathbf{q}} <Q=\sqrt{4\pi n}\). Instead of working with the CS Hamiltonian,
we introduce an equivalent Hamiltonian 
\begin{equation}
\label{equivalent Hamiltonian}
H=\sum _{i}^{N}\frac{1}{2m}(-i\nabla _{i}+e{\mathbf{A}}^{*}({\mathbf{r}}_{i})+{\mathbf{a}}({\mathbf{r}}_{i})+{\mathbf{a}}_{CS}({\mathbf{r}}_{i}))^{2},
\end{equation}
 where \( {\mathbf{a}}({\mathbf{r}}_{i})=-i\hat{z}\times
\widehat{{\mathbf{q}}}a({\mathbf{r}}_{i}) \) is a transverse vector
field. We also define a longitudinal vector field \(
{\mathbf{P}}({\mathbf{r}}_{i})=i\widehat{{\mathbf{q}}}P({\mathbf{r}}_{i})
\) (\( \widehat{{\mathbf{q}}} \) is a unit vector in the \(
{\mathbf{q}} \) direction).  This problem is equivalent to the
original one provided we restrict our attention to states that are
annihilated by the constraints \( \chi ({\mathbf{q}})=a({\mathbf{q}})
\) (\( q<Q \)). We will call states that are annihilated by  these constraints
physical states, that is,  
\begin{equation}
\chi({\mathbf{q}})|\Psi_{phys}>=0.
\end{equation}
We will continue to use notation \( \chi
({\mathbf{q}}) \) for the constraint operator in different
representations.

Using the fields \( a({\mathbf{q}}),P({\mathbf{q}}) \), a unitary
transformation is then constructed which shifts $a$ to absorb the
Fourier components of \( {\mathbf{a}}_{CS}({\mathbf{r}}_{i}) \) for
$q<Q$. In the new representation, whose particles we call composite
particles {(}CP{)}, the Hamiltonian is
\begin{equation}
\label{CP Hamiltonian}
H_{CP}=\sum _{i}^{N}\frac{1}{2m}(-i\nabla _{i}+e{\mathbf{A}}^{*}({\mathbf{r}}_{i})+{\mathbf{a}}({\mathbf{r}}_{i})+2\pi l{\mathbf{P}}({\mathbf{r}}_{i}))^{2}.
\end{equation}
 We neglected all the \( q>Q \) Fourier components of the gauge field in deriving
the above Hamiltonian, implying that our theory will not describe the motion
correctly for large momenta (or short distances). In the following
we will also be using the Random Phase Approximation {(}RPA{)} generalized for
the case of the inhomogeneous densities 
\begin{equation}
\label{RPA}
\sum _{j}e^{i{\mathbf{k}}\cdot {\mathbf{r}}_{j}}\simeq n(2\pi )^{2}\delta ^{2}({\mathbf{k}})+\delta n(-{\mathbf{k}}).
\end{equation}
 The constraints in the CP representation are given by 
\begin{equation}
\label{CP constraints}
\chi ({\mathbf{q}})=-\frac{qa({\mathbf{q}})}{2\pi l}+\rho ({\mathbf{q}})-n({\mathbf{q}}),
\end{equation}
 where \( \rho ({\mathbf{q}})=\sum _{j}^{N}e^{-i{\mathbf{q}}\cdot {\mathbf{r}}_{j}} \)
is the CP density operator and coincides with the corresponding electron operator.

Before proceeding with any further transformations on the Hamiltonian
(\ref{CP Hamiltonian}), we will show how the trial wave function used
to compute the energy of the correlated WC in\cite{yi-fertig} emerges
naturally within this approach\cite{M-S
jain book}.  The crudest approximation for the CP Hamiltonian is
\begin{eqnarray}
 &  & H_{CP}\simeq \sum _{i}^{N}\frac{1}{2m}(-i\nabla _{i}+e{\mathbf{A}}^{*}({\mathbf{r}}_{i}))^{2}\nonumber \\
 &  & +\frac{n}{2m}\sum _{{\mathbf{q}}}^{Q}(a(-{\mathbf{q}})a({\mathbf{q}})+(2\pi l)^{2}P(-{\mathbf{q}})P({\mathbf{q}})).\label{approximate CP Hamiltonian} 
\end{eqnarray}
 In this expression we have assumed that the CP ground state can be
regarded as largely homogeneous, so that the \( \delta n \) is
neglected in the definitions of the RPA (\ref{RPA}) and vector
potential \( {\mathbf{A}}^{*}({\mathbf{r}}_{i}) \), which corresponds
now to the uniform magnetic field. We have also neglected the coupling
between the CPs and the oscillator fields \( a({\mathbf{q}}) \), \(
P({\mathbf{q}}) \). Since this Hamiltonian (\ref{approximate CP
Hamiltonian}) has been artificially made separable into a sum of the
particle and the oscillator terms we can write down the ground state
as a product wave function. The particles are moving in a uniform
magnetic field \( B^{*}=B-2\pi ln/e \), so there is a degeneracy in
this problem, but we assume that their ground state is crystalline
\begin{equation}
\label{CF wave function}
\Psi _{CF}(\{{\mathbf{r}}_{i}\})=\prod _{i}\phi _{{\mathbf{R}}_{i}}({\mathbf{r}}_{i}),
\end{equation}
 where \( \phi _{{\mathbf{R}}_{i}}({\mathbf{r}}_{i}) \) are Gaussians centered
on the lattice sites \( {\mathbf{R}}_{i} \) similar to (\ref{single-particle}),
except that instead of magnetic length \( l_{0} \) there is a new magnetic
length \( l_{0}^{*}=(eB^{*})^{-1/2} \). CF stands for composite fermions. The
oscillator term describes \( N \) independent harmonic oscillators with the
ground state 
\begin{equation}
\label{oscillator wave function}
\Psi _{osc}(\{{\mathbf{q}}\})=\prod _{{\mathbf{q}}}^{Q}e^{-a^{2}({\mathbf{q}})/4\pi l}.
\end{equation}
 Using the constraints (\ref{CP constraints}) we can eliminate the
field degrees of freedom \( a({\mathbf{q}}) \) in favor of the \(
{\mathbf{r}}_{i} \) in the expression for the oscillator wave function
(\ref{oscillator wave function}).  Since the calculation is described
in \cite{M-S jain book} in great detail, we only give the final result
for the projected oscillator wave function
\begin{equation}
\label{projected wave function}
\Psi _{osc}(\{{\mathbf{r}}_{i}\})_{(a({\mathbf{q}})=2\pi l\rho ({\mathbf{q}})/q)}=\prod _{i<j}\left| z_{i}-z_{j}\right| ^{l}e^{-\sum _{j}l\nu |z_{j}|^{2}/4l_{0}^{2}}.
\end{equation}
 Here \( z_{i}=x_i+iy_i \) is the complex coordinate and \( \nu =2\pi
n/eB \) is the filling factor. The approximate CS wave function is a
product of (\ref{CF wave function}) and (\ref{projected wave
function}) and agrees with the equation (2.12) given
in Ref.\cite{yi-fertig}.

However, the fact remains that in this representation the oscillators
and the particles remain strongly coupled. The oscillators are
identified with the magnetoplasmons, which are high-energy degrees of
freedom, while the particles will turn out to form the low-energy
sector. Our next task will be to construct a canonical transformation
to decouple the two sectors, so that we are left with a purely
low-energy theory.

\section{ DECOUPLING AND THE ELECTRON DENSITY OPERATOR IN THE FINAL REPRESENTATION}

Before we turn to the technical details of the decoupling
transformation, it is worthwhile to articulate the philosophy of our
approach. If one were able to find the exact canonical transformation,
and implement it exactly, then one would be left with a final theory
in which the fermions are purely low-energy objects, and the
oscillators are purely high-energy objects. In particular, the
oscillators should obey Kohn's theorem\cite{kohn}, while all reference
to the bare mass should have disappeared from the low-energy fermionic
part of the Hamiltonian. In other words, the bare kinetic energy of
the CFs should be quenched in the final representation. Finally, the
projected electronic density when expressed in the final
representation should obey the magnetic translation algebra
appropriate to the LLL\cite{GMP}. Unfortunately, this program cannot
be implemented fully in practice. What can be implemented is a
sequence of transformations that achieves some measure of the above at
long distance scales (small $q$). We will see that the oscillators do
end up obeying Kohn's theorem, since this is a small-$q$
property. Similarly, the magnetic translation algebra will be seen to
occur in its small-$q$ form. However, while the tendency for the
quenching of the mass will be manifest, the mass depends on all
distance scales, and its quenching cannot be shown within a
long-distance approximation. Our approach will be to {\it assume} the
exact quenching of bare mass, since we know this to be true in the
LLL, and write the final Hamiltonian in the low-energy sector as a
pure interaction term. Thus, while the proximate goal of the canonical
transformation is to decouple the high- and low-energy parts, its
ultimate goal is to obtain the electronic density operator in the
final representation.

We return now to the CP Hamiltonian (\ref{CP Hamiltonian}) and the set of CP
constraints (\ref{CP constraints}). Using the RPA approximation as given by
(\ref{RPA}), the CP Hamiltonian can be recast into the following form 
\begin{eqnarray}
 &  & H_{CP}=\sum _{i}^{N}\frac{{\mathbf{\Pi}} _{j}^{2}}{2m}+\frac{\sqrt{\pi l}}{m}\sum _{{\mathbf{q}}}^{Q}\left( A({\mathbf{q}})\, c^{\dagger }({\mathbf{q}})+\textrm{h}.\textrm{c}.\right) \nonumber \\
 &  & +\frac{\pi l}{m}\sum _{{\mathbf{q}}}^{Q}\sum _{{\mathbf{G}}}^{Q}\left( \left( n\delta _{{\mathbf{G}},0}+\delta n({\mathbf{G}})\right) \right. \nonumber \\
 &  & \times \left. \left( A^{\dagger }({\mathbf{q}})A({\mathbf{q}}-{\mathbf{G}})\hat{q}_{-}\widehat{(q-G)}_{+}\right) +\textrm{h}.\textrm{c}.\right) .\label{RPA Hamiltonian} 
\end{eqnarray}
 The first term in Eq. (\ref{RPA Hamiltonian}) is the CP kinetic energy, with
\( {\mathbf{\Pi}} _{j}=-i\nabla _{j}+e{\mathbf{A}}^{*}({\mathbf{r}}_{j}) \), which only depends
on the particle degrees of freedom. The second is the coupling between the particle
and the auxiliary field (oscillator) degrees of freedom, with \( c({\mathbf{q}})=\hat{q}_{-}\sum _{j}\Pi _{j+}e^{-i{\mathbf{q}}\cdot {\mathbf{r}}_{j}} \)
and the 'destruction' operator \( A({\mathbf{q}})=(a({\mathbf{q}})+i2\pi lP({\mathbf{q}}))/\sqrt{4\pi l} \)\cite{note1}.
The last term describes the oscillators and does not depend on the particle
degrees of freedom. In order to decouple the high energy oscillators from the
low energy CP's we need to compute the commutators between the newly introduced
operators. It is straightforward to deduce from (\ref{a,P commutator}) that
\begin{equation}
\label{A commutator}
\left[ A({\mathbf{q}}),A^{\dagger }({\mathbf{q}}')\right] =(2\pi )^{2}\delta ^{2}({\mathbf{q}}-{\mathbf{q}}').
\end{equation}
 The commutator for the operator \( c({\mathbf{q}}) \) is found from the commutator
of the canonical momenta \( \Pi _{x} \) and \( \Pi _{y} \) and then using
the RPA approximation (\ref{RPA}). The result is

\begin{eqnarray}
\left[ c({\mathbf{q}}),c^{\dagger }({\mathbf{q}}')\right] \simeq \hat{q}_{-}{\hat{q}'}_{+}\left( -4\pi ln\delta n({\mathbf{q}}-{\mathbf{q}}')\right.  &  & \nonumber \\
+\left. 2eB^{*}\left( n(2\pi )^{2}\delta ^{2}({\mathbf{q}}-{\mathbf{q}}')+\delta n({\mathbf{q}}-{\mathbf{q}}')\right) \right) . &  & \label{c commutator} 
\end{eqnarray}
 According to our assumption the average density is a periodic function, as
in (\ref{average density}), therefore the right-hand side of Eq. (\ref{c commutator})
differs from zero only if the difference \( {\mathbf{q}}-{\mathbf{q}}' \) in (\ref{c commutator})
is equal to a reciprocal lattice vector \( {\mathbf{G}} \).

Our task is to decouple the Hamiltonian (\ref{RPA Hamiltonian}) by
eliminating the term that couples the particle and the oscillator
degrees of freedom. We will show how one can construct a canonical
transformation that accomplishes this decoupling. Once the canonical
transformation is found we can derive the electron density operator in
the 'Final' representation {(}FR{)}. Operators in the FR are expressed
in terms of the CF coordinates. We will show below that there are good
reasons to believe that Composite Fermions are the true quasiparticles
in the FQH regime and the FR density operator represents the physical
charge density.

The calculation is a straightforward extension of the procedure given
in reference \cite{M-S jain book} for the case of the homogeneous
liquid, and we relegate the details of this calculation to Appendix
I. In what follows only the results of applying the transformation on
the Hamiltonian (\ref{RPA Hamiltonian}), the density operator and the
set of constraints (\ref{CP constraints}) are presented.

By construction, the term coupling the particles and the oscillators
is not present in the FR Hamiltonian. Substituting Equations
(\ref{A1}), (\ref{c1}) and (\ref{beta}) into the expression for the CF
Hamiltonian (\ref{RPA Hamiltonian}) we find that the oscillator term
in the FR is equal to \( \omega _{c}\sum _{{\mathbf{q}}}A^{\dagger
}({\mathbf{q}})A({\mathbf{q}}) \) with $\omega_c=eB/m$ exactly as in
the liquid, to order \( (\delta n/n)^{2} \). This is a physically
correct result because according to the Kohn's theorem \cite{kohn} the
limit \( \omega _{c}({\mathbf{q}}\rightarrow 0) \) should not depend
on the electron interactions in the lowest Landau level. The kinetic
energy in the FR is

\begin{eqnarray}
T & = & \sum _{j}^{N}\frac{\Pi _{j-}\Pi _{j+}}{2m}+\sum _{j}^{N}\left( \frac{eB^{*}}{2m}-\frac{\pi l}{m}\delta n({\mathbf{r}}_{j})\right) \nonumber \\
 &  & -\frac{1}{2mn}\sum _{{\mathbf{q}}}^{Q}c^{\dagger }({\mathbf{q}})c({\mathbf{q}})\nonumber \\
 &  & +\frac{1}{2mn^{2}}\sum _{{\mathbf{q}}}^{Q}\sum ^{Q}_{{\mathbf{G}}}c^{\dagger }({\mathbf{q}})c({\mathbf{q}}-{\mathbf{G}})\delta n({\mathbf{G}})\hat{q}_{-}\widehat{(q-G)}_{+}.\label{FR kinetic energy} 
\end{eqnarray}
In an ideal calculation the particle kinetic energy should disappear
in the FR; the electronic kinetic energy is subsumed into the
oscillator term. As has been stated above, it is impossible to show
this in a small-$q$ approximation such as the one we are using.

The electron density operator in the FR is obtained by solving the flow equation
that is derived in a way that follows closely the calculation for the kinetic
energy \( T \) leading to the Eq. (\ref{flow equation T}). The result of the
integration of the flow equation is
\begin{eqnarray}
\rho ({\mathbf{q}},\lambda )=\rho ({\mathbf{q}})+\rho _{0}({\mathbf{q}},\lambda )+\frac{q}{4n\sqrt{\pi l(1+\mu ^{2})}}\sum _{{\mathbf{G}}}^{Q}\frac{\delta n({\mathbf{G}})}{N}\left( A({\mathbf{q}}-{\mathbf{G}})\hat{q}_{-}\widehat{(q-G)}_{+}+\textrm{h}.\textrm{c}.\right)  &  & \nonumber \\
+\frac{q(2+\mu ^{2}-2\sqrt{1+\mu ^{2}})}{8\pi ln^{2}\mu ^{4}\sqrt{1+\mu ^{2}}}\sum _{{\mathbf{G}}}^{Q}\frac{\delta n({\mathbf{G}})}{N}\left( c({\mathbf{q}}-{\mathbf{G}})\hat{q}_{-}\widehat{(q-G)}_{+}+\textrm{h}.\textrm{c}.\right) , &  & \label{density1} 
\end{eqnarray}
where \( \mu ^{2}=1/l\nu -1 \). The FR operator \( \rho _{0}({\mathbf{q}},\lambda ) \)
is the leading term in the perturbation expansion of the density in the parameter
\( \delta n/n \) and is formally identical\cite{note3} to the density operator in the case
when the average electron density is uniform\cite{M-S jain book}
\begin{eqnarray}
\rho _{0}({\mathbf{q}},\lambda )=\frac{q}{2\sqrt{\pi l(1+\mu ^{2})}}\left( A({\mathbf{q}})+\textrm{h}.\textrm{c}.\right)  &  & \nonumber \\
-\frac{q\left( \sqrt{1+\mu ^{2}}-1\right) }{4\pi ln\mu ^{2}\sqrt{(1+\mu ^{2})}}\left( c({\mathbf{q}})+\textrm{h}.\textrm{c}.\right) . &  & \label{density0} 
\end{eqnarray}

It is now straightforward to get the FR expression for the set of
constraints (\ref{CP constraints}), just by using the previously
determined FR operators \( A({\mathbf{q}},\lambda ) \) (see Appendix
I, Eq. (\ref{A1})) and \( \rho ({\mathbf{q}},\lambda ) \) (Eqs
(\ref{density1}) and (\ref{density0})). The expression that results is
\begin{eqnarray}
\chi ({\mathbf{q}},\lambda ) & = & \chi _{0}({\mathbf{q}},\lambda )-\delta n({\mathbf{q}})+\frac{q(2+\mu ^{2}-\mu ^{4}-2\sqrt{1+\mu ^{2}})}{8\pi ln^{2}\mu ^{4}\sqrt{1+\mu ^{2}}}\nonumber \\
 &  & \times \sum _{{\mathbf{G}}}^{Q}\delta n({\mathbf{G}})\left( c({\mathbf{q}}-{\mathbf{G}})\hat{q}_{-}\widehat{(q-G)}_{+}+\textrm{h}.\textrm{c}.\right) .\label{constraints1} 
\end{eqnarray}
Here again \( \chi _{0}({\mathbf{q}},\lambda ) \) is the part of the constraint
that corresponds to the case of the uniform average density and is given by
\begin{equation}
\label{constraints0}
\chi _{0}({\mathbf{q}},\lambda )=\rho ({\mathbf{q}})+\frac{q(\sqrt{1+\mu ^{2}}-1)}{4\pi ln\mu ^{2}}\left( c({\mathbf{q}})+\textrm{h}.\textrm{c}.\right) .
\end{equation}
The main observation about Eq. (\ref{constraints1}) is that oscillator
degrees of freedom cancel out up to order \( \delta n/n \), implying
that the constraint acts only on particles. This reassures us of the
self-consistency of the decoupling scheme, since there is no use
decoupling the high and low energy modes in the Hamiltonian if the
constraint still non-trivially couples them.

Because the particles are confined entirely to the lowest Landau
level, one expects the physical charge density operator to obey the
magnetic translation algebra. However this is not true for the density
operator defined by Eq. (\ref{density1}).  We are allowed to modify
the definition of \( \rho ({\mathbf{q}},\lambda ) \) in CP
representation by adding to it any multiple of the constraint, since
in an exact calculation in the physical states the constraint is equal
to zero. Following the same approach as in the liquid states\cite{M-S
jain book} we try the linear combination
\begin{equation}
\label{preferred-combination}
\rho ({\mathbf{q}})-1/(\mu ^{2}+1)\chi ({\mathbf{q}}).
\end{equation}
This operator has the virtue that its FR matrix
elements are of order \( q^{2} \) or higher, consistent with Kohn's theorem\cite{kohn}.
The FR expression of the density operator (\ref{preferred-combination}) (which
we will call the preferred density) is

\begin{eqnarray}
\widetilde{\rho }({\mathbf{q}})=\frac{\mu ^{2}}{\mu ^{2}+1}\rho ({\mathbf{q}})-\frac{q}{4\pi ln(1+\mu ^{2})}\left( c({\mathbf{q}})+\textrm{h}.\textrm{c}.\right) +\frac{\delta n({\mathbf{q}})}{\mu ^{2}+1} &  & \nonumber \\
-\frac{q(1-\sqrt{1+\mu ^{2}})}{4\pi ln(1+\mu ^{2})\mu ^{2}}\sum ^{Q}_{{\mathbf{G}}}\delta n({\mathbf{G}})\left( c({\mathbf{q}}-{\mathbf{G}})\hat{q}_{-}\widehat{(q-G)}_{+}+\textrm{h}.\textrm{c}.\right) . &  & \label{preferred density} 
\end{eqnarray}

The calculation of the commutator 
of the preferred density operators to first order in \( \delta n/n \) gives
\begin{eqnarray}
\left[ \widetilde{\rho }({\mathbf{q}}),\widetilde{\rho }({\mathbf{q}}')\right] =il_{0}^{2}({\mathbf{q}}\times {\mathbf{q}}')\widetilde{\rho }({\mathbf{q}}+{\mathbf{q}}') &  & \nonumber \\
+il_{0}^{2}({\mathbf{q}}\times {\mathbf{q}}')\frac{1}{1+\mu ^{2}}\sum ^{Q}_{{\mathbf{G}}}\frac{\delta n({\mathbf{G}})}{n}\chi _{0}({\mathbf{q}}+{\mathbf{q}}'-{\mathbf{G}},\lambda ). &  & \label{preferred density commutator} 
\end{eqnarray}
Here \( \chi _{0}({\mathbf{q}},\lambda ) \) is the constraint to
zeroth order (in \( \delta n/n \)), Eq. (\ref{constraints0}). We
conclude that the magnetic algebra is satisfied for physical states
that are destroyed by the constraint.  This is a weaker result than
was obtained in the translationally invariant case, but nevertheless
still preserves the equivalence of this theory to the original
electronic theory in the LLL at long distances.

The preferred density encodes many nonperturbative features that are
known to be true for the original electronic problem. It shows the
correct fractional charge of the quasiparticles\cite{laugh}, obeys the
magnetic translation algebra in the small $q$ limit\cite{GMP}, and has
matrix elements of order $q^2$ or higher from the ground state. Thus
it is plausible that simple HF calculations with this density will
capture the essential physics in the FQH regime. Certainly, this
expectation has been borne out in calculations for the liquid
states\cite{sm-work}.

Thus all the features of the translationally invariant Hamiltonian
theory, namely compliance with Kohn's theorem, simultaneous decoupling
of the Hamiltonian and the constraints, and a preferred density that
obeys the algebra of magnetic translations, carry over for the
non-uniform electronic density state.

It is of interest to determine the Hall conductance \( \sigma _{xy} \)
for the non-uniform average density state. In the clean limit, when
there is no external potential,  translation invariance
implies that $\sigma_{xy}=\nu e^2/h$. Our theory does indeed predict
this in the clean limit. In order to see this we need the FR expression
for the current.  Starting with the electron current, and eliminating
the CS vector potential we find the following CP current
\begin{equation}
\label{CP current}
(J_{CP})_{+}({\mathbf{q}})=\frac{1}{m}\sum _{j}\Pi _{j+}e^{-i{\mathbf{q}}\cdot {\mathbf{r}}_{j}}+\frac{2n\sqrt{\pi l}}{m}A({\mathbf{q}})\hat{q}_{+}+\frac{2\sqrt{\pi l}}{m}\sum _{{\mathbf{G}}}\delta n({\mathbf{G}})A({\mathbf{q}}-{\mathbf{G}})\widehat{(q-G)}_{+}.
\end{equation}
The first term in Eq. (\ref{CP current}) is just a definition of the operator
\( c({\mathbf{q}}) \) so to get the FR expression for the current we have to
substitute the FR operators \( A({\mathbf{q}},\lambda ) \) and \( c({\mathbf{q}},\lambda ) \).
The FR current consists of two terms, one that does not contain factors
of \( \delta n \) and another proportional to \( \delta n/n \). The first
term is identical to the FR current for the uniform average density state and
as shown in \cite{M-S jain book} depends only on the oscillator degrees of
freedom. To calculate the first order contribution to the current we use expressions
(\ref{A1}) and (\ref{c1}) that give the first order corrections in \( \delta n/n \)
for the operators \( A({\mathbf{q}},\lambda ) \) and \( c({\mathbf{q}},\lambda ) \)
respectively. We find that it is also independent of the CF coordinates.
Both terms add up to
\begin{equation}
\label{FR current}
J_{+}({\mathbf{q}})=\frac{2n\sqrt{\pi l(1+\mu ^{2})}}{m}A({\mathbf{q}})\hat{q}_{+}+\frac{\sqrt{\pi l(1+\mu ^{2})}}{m}\sum _{{\mathbf{G}}}\delta n({\mathbf{G}})A({\mathbf{q}}-{\mathbf{G}})\widehat{(q-G)}_{+}.
\end{equation}
Because the current in the FR depends only on the operators that
represent the oscillators, we can ignore the particle sector in the
conductance calculation.  The argument in
Ref.\cite{M-S jain book} for the uniform average density case goes through
and gives \( \sigma _{xy}=\nu e^{2}/h \) in the limit \(
{\mathbf{q}}\rightarrow 0 \), which is the correct unquantized Hall
conductance in the clean limit.  Note, however, in the presence of
disorder it is believed that \( \sigma _{xy}\rightarrow 0 \), \(
\sigma _{xx}\rightarrow 0 \) such that \( \rho _{xy} \) is its
classical value. A complete theory including disorder effects is
currently nonexistent, and we will confine ourselves to the clean
limit in the sequel.

\section{HARTREE-FOCK APPROXIMATION}

Having determined the correct canonical transformation by decoupling
the CS Hamiltonian, we shift our focus to the Coulomb interaction that
was hitherto ignored. At the small filling factors for which the
Wigner Crystal occurs, the lowest Landau level approximation is
appropriate. Now the LLL electronic Hamiltonian is given by
\begin{equation}
\label{electron hamiltonian}
H=\frac{1}{2S}\sum _{{\mathbf{q}}}\left( V({\mathbf{q}})\rho ({\mathbf{q}})\rho (-{\mathbf{q}})-V({\mathbf{q}})e^{-q^{2}l_{0}^{2}/2}\rho (0)\right) .
\end{equation}
 Here $S$ is the area of the system and the second term is a result of
un-normal-ordering the original electronic Hamiltonian. This is
necessary since we need to have the full operator $\rho$ in order to
write it in the final representation. The two-dimensional Fourier
transform of the Coulomb potential is suppressed at large momenta \(
{\mathbf{q}} \) by multiplying it with a Gaussian
\begin{equation}
\label{coulomb potential}
V({\mathbf{q}})=\frac{2\pi e^{2}}{q\epsilon }e^{-q^{2}\Lambda ^{2}},
\end{equation}
where a parameter \( \Lambda  \) may be used to interpolate between the pure
Coulomb potential and the Coulomb potential that is effective only when the
the distance between two particles is larger than \( \Lambda  \). \( \epsilon  \)
is the dielectric constant. We emphasize that the form of the Fourier transform
of the potential given in Eq. (\ref{coulomb potential}) does not accurately
describe the effect of the sample thickness, but is rather chosen for illustrative
purposes, since it is computationally convenient. The second term in Eq. (\ref{electron hamiltonian})
is a constant that will be ignored in what follows. The first term of the electronic
Coulomb interaction when transformed to the CF coordinates will serve as our
model Hamiltonian
\begin{equation}
\label{Hamiltonian0}
H=\frac{1}{2S}\sum _{{\mathbf{q}}}V({\mathbf{q}})\widetilde{\rho }({\mathbf{q}})\widetilde{\rho }(-{\mathbf{q}}).
\end{equation}
 The density operator \( \widetilde{\rho }({\mathbf{q}}) \) is given by Eq.
(\ref{preferred density}). It is useful at this point to rewrite it so that
the dependence on the modulated average density \( \delta n \) is explicit\cite{note4},
\begin{eqnarray}
\widetilde{\rho }({\mathbf{q}}) & = & (1-c^{2})\rho ({\mathbf{q}})-il_{0}^{2}\sum _{j}{\mathbf{q}}\times {\mathbf{\Pi}} ^{*}_{j}e^{-i{\mathbf{q}}\cdot {\mathbf{r}}_{j}}\nonumber \\
 & - & c^{2}\sum _{{\mathbf{G}},{\mathbf{G}}\neq {\mathbf{q}}}\frac{\delta n({\mathbf{G}})}{N}\frac{{\mathbf{q}}\cdot {\mathbf{G}}}{{\mathbf{G}}^{2}}e^{i({\mathbf{G}}-{\mathbf{q}})\cdot {\mathbf{r}}_{j}}\nonumber \\
 & + & \frac{il_{0}^{2}c}{c+1}\sum _{{\mathbf{G}}}\frac{\delta n({\mathbf{G}})}{N}\sum _{j}{\mathbf{q}}\times {\mathbf{\Pi}} _{j}^{*}e^{i({\mathbf{G}}-{\mathbf{q}})\cdot {\mathbf{r}}_{j}}.\label{preferred density1} 
\end{eqnarray}
Here \( {\mathbf{\Pi}} ^{*}_{j} \) is the momentum operator that corresponds to the uniform
average density case (\( {\mathbf{\Pi}} ^{*}_{j}={\mathbf{\Pi}} _{j}(\delta n=0) \)) and \( c=\sqrt{l\nu } \).
It will also be convenient to have separate symbols for the different orders
of the \( \delta n \) in Eq. (\ref{preferred density1}), so we write \( \widetilde{\rho }({\mathbf{q}})=\widetilde{\rho }_{0}({\mathbf{q}})+\sum _{{\mathbf{G}}}\delta n({\mathbf{G}})\widetilde{\rho }_{1}({\mathbf{q}},{\mathbf{G}}) \).

The Hamiltonian (\ref{Hamiltonian0}) describes a many-body CF problem
that we will treat within the Hartree-Fock approximation. We
justify the use of the HF approximation by arguing that the CF is the
true quasiparticle in the FQH regime. It will be
assumed throughout this study that the average density modulation is
small compared to the uniform background (\( \delta n/n\ll 1 \)) so as
a convenient basis we will choose the wave-functions of the free CF
moving in the uniform magnetic field \( B^{*}=B-2\pi ln/e \). The
Landau gauge will be used in what follows. Wavefunctions will be
denoted as \( \left| n,X\right\rangle \), with \( n \) as a CF Landau
level index and \( X \) as a kinetic momentum component in the \( y \)
direction.

We will now derive the HF Hamiltonian. The model Hamiltonian (\ref{Hamiltonian0})
in the \( \left| n,X\right\rangle  \) basis may be written
\begin{eqnarray}
H & = & \frac{1}{2S}\sum _{{\mathbf{q}}}V({\mathbf{q}})\sum _{n_{1}X_{1},\ldots ,n_{3}X_{3}}\left\langle n_{1}X_{1}\right| \widetilde{\rho }({\mathbf{q}})\left| n_{2}X_{2}\right\rangle \nonumber \\
 &  & \times \left\langle n_{2}X_{2}\right| \widetilde{\rho }(-{\mathbf{q}})\left| n_{3}X_{3}\right\rangle c^{\dagger }_{n_{1},X_{1}}c_{n_{3},X_{3}}\nonumber \\
 &  & +\frac{1}{2S}\sum _{{\mathbf{q}}}V({\mathbf{q}})\sum _{n_{1}X_{1},\ldots ,n_{4}X_{4}}\left\langle n_{1}X_{1}\right| \widetilde{\rho }({\mathbf{q}})\left| n_{4}X_{4}\right\rangle \nonumber  \\
 &  & \times \left\langle n_{2}X_{2}\right| \widetilde{\rho }(-{\mathbf{q}})\left| n_{3}X_{3}\right\rangle c^{\dagger }_{n_{1},X_{1}}c^{\dagger }_{n_{2},X_{2}}c_{n_{3}X_{3}}c_{n_{4}X_{4}},\label{Hamiltonian1} 
\end{eqnarray}
where \( c^{\dagger }_{n,X} \) (\( c_{n,X} \)) is the CF creation (destruction)
operator. The usual HF pairings are made in the two-body term of the Hamiltonian
(\ref{Hamiltonian1}), giving two contributions - a direct and an exchange term.
Because the \( X_{i} \) dependence of the density matrix elements in Eq. (\ref{Hamiltonian1})
is very simple
\begin{eqnarray}
\left\langle n_{1}X_{1}\right| \widetilde{\rho }_{0}({\mathbf{q}})\left| n_{2}X_{2}\right\rangle  & = & \left\langle n_{1}\right| \widetilde{\rho }_{0}({\mathbf{q}})\left| n_{2}\right\rangle e^{-iq_{x}(X_{1}+X_{2})/2}\delta _{X_{1},X_{2}-q_{y}l_{0}^{*2}}\label{0matrix element} \\
\left\langle n_{1}X_{1}\right| \widetilde{\rho }_{1}({\mathbf{q}},{\mathbf{G}})\left| n_{2}X_{2}\right\rangle  & = & \left\langle n_{1}\right| \widetilde{\rho }_{1}({\mathbf{q}},{\mathbf{G}})\left| n_{2}\right\rangle \nonumber \\
 &  & \times e^{-i(q_{x}-G_{x})(X_{1}+X_{2})/2}\delta _{X_{1},X_{2}-(q_{y}-G_{y})l_{0}^{*2}}\label{1matrix element} 
\end{eqnarray}
(here \( l^{*}_{0}=(eB^{*})^{-1/2} \) is the CF magnetic length) one can eliminate
all the dependence on the momentum \( X_{i} \) in the Eq. (\ref{Hamiltonian1}).
To this end we introduce the operator 
\begin{equation}
\label{order parameter}
\Delta _{nn'}({\mathbf{q}})=\frac{1}{g}\sum _{X}e^{-iq_{x}X-iq_{x}q_{y}l_{0}^{*2}/2}c^{\dagger }_{n,X}c_{n',X+q_{y}l_{0}^{*2}},
\end{equation}
 where \( g \) is the degeneracy of a Landau level. \( \Delta _{nn'}({\mathbf{G}}) \)
is the order parameter of the density modulation corresponding to the wave-vector
\( {\mathbf{G}} \). Note that \( \Delta _{nn'}({\mathbf{q}}) \) has the property
\begin{equation}
\label{sum_rule1}
\sum _{n}\Delta _{nn}(0)=\nu .
\end{equation}
 After doing sums over \( X_{i} \) we find the following contributions to the
HF Hamiltonian: 

\begin{enumerate}
\item A one-body term, zeroth order in \( \delta n \):
\begin{equation}
\label{HF1}
H^{00}_{ob}=\frac{g}{2S}\sum _{{\mathbf{q}}}V({\mathbf{q}})\sum _{n_{1},n_{2},n_{3}}\left\langle n_{1}\right| \widetilde{\rho }_{0}({\mathbf{q}})\left| n_{2}\right\rangle \left\langle n_{2}\right| \widetilde{\rho }_{0}(-{\mathbf{q}})\left| n_{3}\right\rangle \Delta _{n_{1}n_{3}}(0).
\end{equation}
 
\item A one-body term, first order in \( \delta n \):
\begin{eqnarray}
H^{01}_{ob} & = & \frac{g}{2S}\sum _{{\mathbf{q}},{\mathbf{G}}}V({\mathbf{q}})\sum _{n_{1},n_{2},n_{3}}\left( \left\langle n_{1}\right| \widetilde{\rho }_{1}(-{\mathbf{q}},{\mathbf{G}})\left| n_{2}\right\rangle \left\langle n_{2}\right| \widetilde{\rho }_{0}({\mathbf{q}})\left| n_{3}\right\rangle \right. \nonumber \\
 &  & \left. +\left\langle n_{1}\right| \widetilde{\rho }_{0}(-{\mathbf{q}})\left| n_{2}\right\rangle \left\langle n_{2}\right| \widetilde{\rho }_{1}({\mathbf{q}},{\mathbf{G}})\left| n_{3}\right\rangle \right) \nonumber \\
 &  & \times \delta n({\mathbf{G}})e^{il_{0}^{*2}{\mathbf{q}}\times {\mathbf{G}}/2}\Delta _{n_{1}n_{3}}(-{\mathbf{G}}).\label{HF2} 
\end{eqnarray}

\item A one-body term, second order in \( \delta n \):
\begin{eqnarray}
H^{11}_{ob} & = & \frac{g}{2S}\sum _{{\mathbf{q}},{\mathbf{G}},{\mathbf{G}}_{1}}V({\mathbf{q}})\sum _{n_{1},n_{2},n_{3}}\left\langle n_{1}\right| \widetilde{\rho }_{1}(-{\mathbf{q}},{\mathbf{G}})\left| n_{2}\right\rangle \left\langle n_{2}\right| \widetilde{\rho }_{1}({\mathbf{q}},{\mathbf{G}}_{1})\left| n_{3}\right\rangle \nonumber \\
 &  & \times \delta n({\mathbf{G}})\delta n({\mathbf{G}}_{1})e^{il_{0}^{*2}\left( {\mathbf{q}}\times ({\mathbf{G}}+{\mathbf{G}}_{1})-{\mathbf{G}}\times {\mathbf{G}}_{1}\right) /2}\Delta _{n_{1}n_{3}}(-{\mathbf{G}}-{\mathbf{G}}_{1}).\label{HF3} 
\end{eqnarray}

\item A two-body term, zeroth order in \( \delta n \),  direct and  exchange
contributions:
\begin{eqnarray}
H^{00}_{tb} & = & \frac{g^{2}}{S}\sum _{{\mathbf{G}}}V({\mathbf{G}})\sum _{n_{1},\ldots ,n_{4}}\left\langle n_{1}\right| \widetilde{\rho }_{0}(-{\mathbf{G}})\left| n_{4}\right\rangle \left\langle n_{2}\right| \widetilde{\rho }_{0}({\mathbf{G}})\left| n_{3}\right\rangle \nonumber \\
 &  & \times \left\langle \Delta _{n_{1}n_{4}}(-{\mathbf{G}})\right\rangle \Delta _{n_{2}n_{3}}({\mathbf{G}})-\sum _{{\mathbf{q}},{\mathbf{G}}}V({\mathbf{q}})\sum _{n_{1},\ldots ,n_{4}}\left\langle n_{1}\right| \widetilde{\rho }_{0}(-{\mathbf{q}})\left| n_{4}\right\rangle \nonumber \\
 &  & \times \frac{g}{S}\left\langle n_{2}\right| \widetilde{\rho }_{0}({\mathbf{q}})\left| n_{3}\right\rangle \left\langle \Delta _{n_{1}n_{3}}({\mathbf{G}})\right\rangle \Delta _{n_{2}n_{4}}(-{\mathbf{G}})e^{il_{0}^{*2}{\mathbf{G}}\times {\mathbf{q}}}.\label{HF4} 
\end{eqnarray}

\item A two-body term, first order in \( \delta n \),  direct contributions:
\begin{eqnarray}
H^{01}_{tbd} & = & \frac{g^{2}}{S}\sum _{{\mathbf{G}},{\mathbf{G}}_{1}}\sum _{n_{1},\ldots ,n_{4}}\left( V(-{\mathbf{G}}-{\mathbf{G}}_{1})\left\langle n_{1}\right| \widetilde{\rho }_{1}({\mathbf{G}}+{\mathbf{G}}_{1},{\mathbf{G}})\left| n_{4}\right\rangle \right. \nonumber \\
 &  & \times \left\langle n_{2}\right| \widetilde{\rho }_{0}(-{\mathbf{G}}-{\mathbf{G}}_{1})\left| n_{3}\right\rangle +V(-{\mathbf{G}}_{1})\left\langle n_{1}\right| \widetilde{\rho }_{0}({\mathbf{G}}_{1})\left| n_{4}\right\rangle \nonumber \\
 &  & \times \left. \left\langle n_{2}\right| \widetilde{\rho }_{1}(-{\mathbf{G}}_{1},{\mathbf{G}})\left| n_{3}\right\rangle \right) \delta n({\mathbf{G}})\left\langle \Delta _{n_{1}n_{4}}({\mathbf{G}}_{1})\right\rangle \Delta _{n_{2}n_{3}}(-{\mathbf{G}}-{\mathbf{G}}_{1}).\label{HF5} 
\end{eqnarray}

\item A two-body term, first order in \( \delta n \),  exchange contributions:
\begin{eqnarray}
H^{01}_{tbe} & = & -\frac{g}{S}\sum _{{\mathbf{q}},{\mathbf{G}},{\mathbf{G}}_{1}}\sum _{n_{1},\ldots ,n_{4}}V({\mathbf{q}})\left( \left\langle n_{1}\right| \widetilde{\rho }_{1}(-{\mathbf{q}},{\mathbf{G}})\left| n_{4}\right\rangle \left\langle n_{2}\right| \widetilde{\rho }_{0}({\mathbf{q}})\left| n_{3}\right\rangle \right. \nonumber \\
 &  & \times e^{il_{0}^{*2}\left( {\mathbf{G}}\times ({\mathbf{q}}-{\mathbf{G}}_{1})/2+{\mathbf{G}}_{1}\times {\mathbf{q}}\right) }+\left\langle n_{1}\right| \widetilde{\rho }_{0}(-{\mathbf{q}})\left| n_{4}\right\rangle \nonumber \\
 &  & \times \left. \left\langle n_{2}\right| \widetilde{\rho }_{1}({\mathbf{q}},{\mathbf{G}})\left| n_{3}\right\rangle e^{il_{0}^{*2}\left( {\mathbf{G}}\times ({\mathbf{q}}+{\mathbf{G}}_{1})/2+{\mathbf{G}}_{1}\times {\mathbf{q}}\right) }\right) \nonumber \\
 &  & \times \delta n({\mathbf{G}})\left\langle \Delta _{n_{1}n_{3}}({\mathbf{G}}_{1})\right\rangle \Delta _{n_{2}n_{4}}(-{\mathbf{G}}-{\mathbf{G}}_{1}).\label{HF6} 
\end{eqnarray}
 
\item A two-body term, second order in \( \delta n \),  direct contribution:
\begin{eqnarray}
H^{11}_{tbd} & = & \frac{g^{2}}{S}\sum _{{\mathbf{G}},{\mathbf{G}}_{1},{\mathbf{G}}_{2}}\sum _{n_{1},\ldots ,n_{4}}V(-{\mathbf{G}}-{\mathbf{G}}_{1})\nonumber \\
 &  & \times \left\langle n_{1}\right| \widetilde{\rho }_{1}({\mathbf{G}}+{\mathbf{G}}_{1},{\mathbf{G}}_{1})\left| n_{4}\right\rangle \left\langle n_{2}\right| \widetilde{\rho }_{1}(-{\mathbf{G}}-{\mathbf{G}}_{1},{\mathbf{G}}_{2})\left| n_{3}\right\rangle \nonumber \\
 &  & \times \delta n({\mathbf{G}}_{1})\delta n({\mathbf{G}}_{2})\left\langle \Delta _{n_{1}n_{4}}({\mathbf{G}})\right\rangle \Delta _{n_{2}n_{3}}(-{\mathbf{G}}-{\mathbf{G}}_{1}-{\mathbf{G}}_{2}).\label{HF7} 
\end{eqnarray}

\item A two-body term, second order in \( \delta n \),  exchange contribution:
\begin{eqnarray}
H^{11}_{tbe} & = & -\frac{g}{S}\sum _{{\mathbf{q}},{\mathbf{G}},{\mathbf{G}}_{1},{\mathbf{G}}_{2}}V({\mathbf{q}})\sum _{n_{1},\ldots ,n_{4}}\left\langle n_{1}\right| \widetilde{\rho }_{1}(-{\mathbf{q}},{\mathbf{G}}_{1})\left| n_{4}\right\rangle \nonumber \\
 &  & \times \left\langle n_{2}\right| \widetilde{\rho }_{1}({\mathbf{q}},{\mathbf{G}}_{2})\left| n_{3}\right\rangle \left\langle \Delta _{n_{1}n_{3}}({\mathbf{G}})\right\rangle \delta n({\mathbf{G}}_{1})\delta n({\mathbf{G}}_{2})\Delta _{n_{2}n_{4}}(-{\mathbf{G}}-{\mathbf{G}}_{1}-{\mathbf{G}}_{2})\nonumber \\
 &  & \times e^{il_{0}^{*2}\left( ({\mathbf{G}}_{1}+{\mathbf{G}}_{2})\times {\mathbf{q}}+{\mathbf{G}}\times ({\mathbf{G}}_{1}-{\mathbf{G}}_{2})-{\mathbf{G}}_{1}\times {\mathbf{G}}_{2}+2{\mathbf{G}}\times {\mathbf{q}}\right) /2}.\label{HF8} 
\end{eqnarray}
 
\end{enumerate}
The matrix elements of the density operator can be calculated using the formulas
(\ref{matrix element1}), (\ref{matrix element2}) and (\ref{matrix element3})
that are given in the Appendix II. The momenta \( {\mathbf{G}}_{i} \) run over
a discrete set of reciprocal lattice vectors. The momentum \( {\mathbf{q}} \)
is a continuous variable. The summation over \( {\mathbf{q}} \) in those terms
of the Hamiltonian where it appears can be done in a closed form as we show
in Appendix III, because the potential in Eq. (\ref{coulomb potential}) was
chosen so that these integrals could be performed analytically.

We will group all the non-operator entries in the Equations (\ref{HF1})-(\ref{HF8})
under the notation \( U_{n_{1}n_{2}}({\mathbf{G}}) \) (renaming the dummy summation
variables where necessary) and represent the HF Hamiltonian in a form convenient
for further discussion,
\begin{equation}
\label{HF Hamiltonian}
H_{HF}=g\sum _{{\mathbf{G}},n_{1},n_{2}}U_{n_{1}n_{2}}({\mathbf{G}})\Delta _{n_{1}n_{2}}({\mathbf{G}}).
\end{equation}
Obviously \( U_{n_{1}n_{2}}({\mathbf{G}}) \) depends on the expectation value
of the order parameter operator \( \Delta _{n_{1}n_{2}}({\mathbf{G}}) \) both
directly and through the density modulation \( \delta n \), because
\begin{eqnarray}
\delta n({\mathbf{G}}) & = & g\sum _{n_{1},n_{2}}\left\langle n_{1}\right| \widetilde{\rho }_{0}({\mathbf{G}})\left| n_{2}\right\rangle \left\langle \Delta _{n_{1}n_{2}}({\mathbf{G}})\right\rangle \nonumber \\
 &  & +g\sum _{{\mathbf{G}}_{1},n_{1},n_{2}}\left\langle n_{1}\right| \widetilde{\rho }_{1}({\mathbf{G}},{\mathbf{G}}_{1})\left| n_{2}\right\rangle \delta n({\mathbf{G}}_{1})\left\langle \Delta _{n_{1}n_{2}}({\mathbf{G}}-{\mathbf{G}}_{1})\right\rangle .\label{delta n} 
\end{eqnarray}
 In the \( i \)'th iteration of the numerical HF procedure \( \Delta _{n_{1}n_{2}}({\mathbf{G}}) \)
is calculated using the solution of the \( i-1 \)'st (previous) iteration.
The density modulation is then calculated as a numerical solution the system
of the linear equations defined in Eq. (\ref{delta n}).

Having found the HF Hamiltonian, we can solve it to find the single-particle
spectrum of the many-body system. We will assume that the CF form a Wigner lattice
with one particle per unit cell. The reciprocal lattice constant for a triangular 
lattice is given by
\begin{equation}
\label{G_0}
G_{0}=\frac{1}{l_{0}}\sqrt{\frac{4\pi \nu }{\sqrt{3}}}.
\end{equation}
However, we will find that in some regions of filling factor the
triangular lattice is not the ground state, and we will explore other
lattice structures. 

We have used two different schemes to perform the
calculation, one due to C\^{o}t\'{e} and
MacDonald\cite{cote-macdonald} and the other due to Yoshioka and
Lee\cite{yoshioka-lee}.  Below we will outline the essence of each of
these methods.

The method by C\^{o}t\'{e} and MacDonald (CM) starts from the
single-particle Green's function which they define as
\begin{equation}
\label{green's_function}
G_{n_{1}n_{2}}(X_{1},X_{2},\tau )=-\left\langle Tc_{n_{1},X_{1}}(\tau )c^{\dagger }_{n_{2},X_{2}}(0)\right\rangle ,
\end{equation}
here \( T \) is the time-ordering operator. The relationship of the Green's
function Fourier transform to the physically relevant expectation value of the
order parameter is 
\begin{eqnarray}
\left\langle \Delta _{n_{1}n_{2}}({\mathbf{G}})\right\rangle =G_{n_{2}n_{1}}({\mathbf{G}},\tau =0^{-}) &  & \nonumber \\
\equiv \frac{1}{g}\sum _{X_{1},X_{2}}G_{n_{2}n_{1}}(X_{2},X_{1},\tau =0^{-})e^{-iG_{x}(X_{1}+X_{2})/2}\delta _{X_{1},X_{2}-G_{y}l^{*2}_{0}}. &  & \label{order_parameter1} 
\end{eqnarray}
The set of crystal order parameters \( \left\langle \Delta _{n_{1}n_{2}}({\mathbf{G}})\right\rangle  \)
is then used to find the ground-state energy 
\begin{equation}
\label{ground-state}
E_{HF}=\frac{\epsilon _{1}}{\nu }\sum _{{\mathbf{G}},n_{1},n_{2}}U_{n_{1}n_{2}}({\mathbf{G}})\left\langle \Delta _{n_{1}n_{2}}({\mathbf{G}})\right\rangle ,
\end{equation}
where \( \epsilon _{1}=1 \) if it multiplies those terms of \(
U_{n_{1}n_{2}}({\mathbf{G}}) \) that are given by the Equations
(\ref{HF1})-(\ref{HF3}) (one-body terms) and \( \epsilon _{1}=1/2 \)
if it multiplies terms that are given by the Equations
(\ref{HF4})-(\ref{HF8}) (two-body terms). The excitation gap \( E_{g}
\), also called activation energy, can be deduced from the chemical
energy and the single-energy density of states \( d(E) \) which is
related to the Green's function through
\begin{equation}
\label{density_of_states}
d(E)=-\frac{1}{\pi }\sum _{n}\Im G_{nn}({\mathbf{G}}=0,i\omega _{j}\rightarrow E+i\delta ),
\end{equation}
 here \( \Im G_{nn} \) is the imaginary part of the operator \( G_{nn}
\), \( \omega _{j} \) are the Matsubara frequencies, \( \delta \) is a
small smoothing parameter.

All of the above is predicated on knowing the Green's function. We
derive the Green's function equation of motion in the usual way by
taking the commutator of the Hamiltonian (\ref{HF Hamiltonian}) with a
single particle destruction operator \( c_{nX} \)
\begin{eqnarray}
\left( i\omega _{j}+\frac{\mu }{\hbar }\right) G_{n_{1}n_{2}}({\mathbf{G}},\omega _{j}) &  & \nonumber \\
-\sum _{{\mathbf{G}},n_{3}}\frac{1}{\hbar }U_{n_{1}n_{3}}({\mathbf{G}}_{1}-{\mathbf{G}})G_{n_{3}n_{2}}({\mathbf{G}}_{1},\omega _{j})e^{i{\mathbf{G}}\times {\mathbf{G}}_{1}l^{*2}_{0}} & =\delta _{n_{1},n_{2}}\delta _{{\mathbf{G}},0}, & \label{equation_motion} 
\end{eqnarray}
where \( \mu  \) is the the chemical potential. The system of Equations (\ref{equation_motion})
is solved for the Green's function by diagonalizing its left-hand side with
respect to the indices \( n_{3} \) and \( {\mathbf{G}}_{1} \). One can find
the expectation value of the order parameter and the density of states
once the chemical potential is known. The chemical potential in turn is calculated by filling up the correct number of states, that is, by  using  Eq. (\ref{sum_rule1}). 

The numerical iterative scheme starts by assuming a Gaussian form for the order
parameters. (The exact expression depends on the filling factor and the state
that is being constructed and will be discussed later). This initial set of
\( \left\langle \Delta _{n_{1}n_{2}}({\mathbf{G}})\right\rangle  \) is then used
to compute the effective potential \( U_{n_{1}n_{2}}({\mathbf{G}}) \). Next the
equation of motion is solved to get a new set of order parameters and the process
is repeated until the \( \left\langle \Delta _{n_{1}n_{2}}({\mathbf{G}})\right\rangle  \)
converge with some prescribed accuracy. Another way to check the accuracy of
the numerical solution is by using the following useful sum rule\cite{cote-macdonald}
\begin{equation}
\label{sum-rule2}
\sum _{{\mathbf{G}},n_{1},n_{2}}\left\langle \Delta _{n_{1}n_{2}}({\mathbf{G}})\right\rangle ^{2}=\nu .
\end{equation}

The idea behind the second method, that of Yoshioka and Lee
\cite{yoshioka-lee} (YL), is to diagonalize the one-body HF Hamiltonian
that can be rewritten in terms of the CF creation and destruction
operators as

\begin{equation}
\label{HF-YL1}
H=\sum _{{\mathbf{G}},n_{1},n_{2},X}U_{n_{1}n_{2}}({\mathbf{G}})e^{-iG_{x}X}c^{\dagger }_{n_{1},X-G_{y}l^{*2}_{0}/2}\, c_{n_{2},X+G_{y}l_{0}^{*2}/2}.
\end{equation}
We assume that the CF form a periodic lattice with primitive translation vectors
of the reciprocal lattice that are given by \( {\mathbf{Q}}_{1}=(Q_{0},0) \)
and \( {\mathbf{Q}}_{2}=Q_{0}(p/q,\alpha ) \). The first unitary transformation
on the Hamiltonian (\ref{HF-YL1}) is defined by
\begin{equation}
\label{TR-YL1}
a_{n_{1},X,Y}=\frac{1}{\sqrt{s_{m}}}\sum ^{s_{m}}_{s=0}e^{-is\alpha Q_{0}Y}c_{n_{1},X+s\alpha Q_{0}l^{*2}_{0}},
\end{equation}
here \( s_{m}=L/\alpha Q_{0}l^{*2}_{0} \), \( L \) is the linear dimension
of the system, \( 0\leq X\leq \alpha Q_{0}l^{*2}_{0} \) and \( 0\leq Y<2\pi /\alpha Q_{0} \).
After making the transformation (\ref{TR-YL1}) the Hamiltonian is
\begin{equation}
\label{HF-YL2}
H=\sum _{{\mathbf{G}},n_{1},n_{2},X,Y}U_{n_{1}n_{2}}({\mathbf{G}})e^{-iG_{x}X+iG_{y}Y+iG_{x}G_{y}l^{*2}_{0}/2}a^{\dagger }_{n_{1},X,Y}\, a_{n_{2},X,Y+G_{x}l_{0}^{*2}}.
\end{equation}
The variable \( Y \) in Eq. (\ref{HF-YL2}) is coupled through \( G_{x}l_{0}^{*2} \).
If this number is commensurate with \( 2\pi /\alpha Q_{0} \), which is the
period of the variable \( Y \), then we can simplify the Hamiltonian even further.
Suppose then that the parameters are such that \( NQ_{0}l^{*2}_{0}/q=M2\pi /\alpha Q_{0} \),
with \( M \) and \( N \) integers. We then introduce a new operator
\begin{equation}
\label{TR-YL2}
b_{n,j,X,Y}=a_{n,X,Y+jQ_{0}l^{*2}_{0}/q},
\end{equation}
 where \( 1\leq j\leq N \) and \( 0\leq Y<l^{*2}_{0}Q_{0}/qM \). After inserting
Eq. (\ref{TR-YL2}) into (\ref{HF-YL2}) the expression for the Hamiltonian
is
\begin{eqnarray}
H & = & \sum _{{\mathbf{G}}}\sum _{X,Y}\sum _{n_{1},j,n_{2},k}U_{n_{1}n_{2}}({\mathbf{G}})e^{-iG_{x}X+iG_{y}Y}\nonumber \\
 &  & \times e^{iQ_{0}Q_{y}jl^{*2}_{0}/q+iG_{x}G_{y}l^{*2}_{0}/2}b^{\dagger }_{n_{1},j,X,Y}\, b_{n_{2},k,X,Y}\delta _{k,j+Q_{x}q/Q_{0}}.\label{HF-YL3} 
\end{eqnarray}
For every pair \( (X,Y) \) that takes values in the rectangular domain defined
earlier the Hamiltonian (\ref{HF-YL3}) can be diagonalized in the indices \( n_{1},j \)
and \( n_{2},k \). The single-particle energies \( E_{n_{1},j}(X,Y) \) that
result thereby are continuous in the variables \( X,Y \) and form energy bands.
There are \( n_{m}N \) energy bands (where \( 1\leq n_{1},n_{2}\leq n_{m} \))
and there is a large energy gap between the lower \( Mq \)'th and \( Mq+1 \)'st
bands. The CF state that has the chemical potential in this large gap should
have the lowest energy. Such a state occurs when the CF filling factor is \( \nu ^{*}=qM/N \),
where we have defined the CF filling factor as \( \nu ^{*}=2\pi l^{*2}_{0}n \).
It is easy to derive the expression for the expectation value of the order parameter,
by applying the transformations (\ref{TR-YL1}) and (\ref{TR-YL2}) to the definition
(\ref{order parameter}) to get
\begin{eqnarray}
\left\langle \Delta _{n_{1}n_{2}}({\mathbf{G}})\right\rangle  & = & \frac{2\pi l^{*2}_{0}}{L^{2}}\sum _{X,Y}\sum _{j,k}e^{-iG_{x}X+iG_{y}Y}\nonumber \\
 &  & \times e^{iQ_{0}Q_{y}jl^{*2}_{0}/q+iG_{x}G_{y}l^{*2}_{0}/2}\left\langle b^{\dagger }_{n_{1},j,X,Y}\, b_{n_{2},k,X,Y}\right\rangle \delta _{k,j+Q_{x}q/Q_{0}}.\label{order-parameter-YL} 
\end{eqnarray}
As in the case of the previous method we find a solution to the HF problem by
iterating until the order parameters converge. We calculate the ground state
energy using Eq. (\ref{ground-state}) and the excitation gap \( E_{g} \) as
a smallest separation between the \( Mq \)'th and \( Mq+1 \)'st bands\cite{note5}.

While the  method of CM is numerically efficient
it is sometimes difficult to extract the the excitation gap from the smoothed
density of states. There is no uncertainty in determining \( E_{g} \) when
the  method of YL is used.

\section{RESULTS}

Our experimental motivation is the work by Jiang et al.
\cite{jiang1,jiang2} where the transport properties were measured
around \( \nu =1/5 \) Landau level filling. In \cite{jiang1} an
insulating phase was identified just above \( \nu =1/5 \) at \( \nu
=0.21 \) by observing a large peak of the longitudinal resistance \(
R_{xx} \) as a function of the external magnetic field. The activation
energy was estimated from the Arrhenius plot at \( E_{g}\sim 0.63 \) K
(with \( B\approx 20 \) T). The striking thing is that the magnitude
of the activation gap compares very poorly with the results obtained
from HF for the usual electron solid.  The excitation energies for the
triangular \emph{electron} lattice with one particle per unit cell are
given in Table 1. We use the modified Coulomb potential given by
(\ref{coulomb potential}) and present results for different values of
the thickness parameter \( \Lambda \). The calculation was done in the
lowest Landau level approximation and for \( \Lambda =0 \) it
reproduces previous results\cite{yoshioka-lee}. The energies are given
in units of \( e^{2}/\epsilon l_{0} \). In the same units the
experimental result is \( E_{g}\sim 2.8\times 10^{-3}e^{2}/\epsilon
l_{0} \), at least two orders of magnitude smaller than the theory.

Table 1 {Electron WC ground state and activation energies for different 
values of \( \Lambda \).}

\vspace{0.3cm}
{\centering \begin{tabular}{|c|c|c|c|r|}
\hline 
\( \Lambda  \)&
\( 0 \)&
\( l_{0}/2 \) &
\( l_{0} \)&
\( 3l_{0}/2 \)\\
\hline 
\hline 
\( E_{HF} \)&
\multicolumn{1}{|r|}{\( -0.3220 \)}&
\multicolumn{1}{|r|}{\( -0.3137 \)}&
\multicolumn{1}{|r|}{\( -0.2859 \)}&
\multicolumn{1}{|r|}{\( -0.2413 \)}\\
\hline 
\( E_{g}(e^{2}/\epsilon l_{0}) \)&
\multicolumn{1}{|r|}{\( 0.4728 \)}&
\multicolumn{1}{|r|}{\( 0.5080 \)}&
\multicolumn{1}{|r|}{\( 0.5080 \)}&
\( 0.4468 \)\\
\hline 
\end{tabular}\par}
\vspace{0.3cm}

We expect some reduction in the value of \( E_{g} \) when the relaxation of
the lattice is accounted for\cite{cockayne,FHM}, but it is difficult to believe
that this correction would nearly exactly cancel the unrelaxed excitation energy.
Besides one would not expect the \( E_{g}(\nu ) \) for the electron WC to be
non-monotonic as observed experimentally\cite{jiang2}. 

Now we proceed to carry out our program of considering crystals of CFs.

\subsection{Crystals of Composite Fermions with two vortices attached}

Let us examine how well CFs with \( l=2 \) describe the experimental
situation. An electronic filling factor of $\nu$ corresponds to a CF
filling factor \( \nu ^{*}=\nu /(1-l\nu )=1/3 \).  So the lowest CF
Landau level is partially filled and it is reasonable to expect that
the composite fermions form a lattice. As in the electron solid
calculation only the lowest CF Landau level is kept (\( n_{m}=1
\)). Keeping two CF Landau levels (\( n_{m}=2 \)) we find similar
results, indicating that including more Landau levels does not
influence the calculation. Because the CF and electron effective
potential \( U_{00}({\mathbf{q}}) \) in Eq. (\ref{HF Hamiltonian}) are
different momentum functions it is not obvious that the CF lattice is
triangular as is the case for the electron lattice. The functional
form of \( U_{00}({\mathbf{q}}) \) may be suggestive in that
respect. One expects it to have a minimum at about the momentum \(
{\mathbf{q}} \) equal to the shortest reciprocal vector. This argument
cannot be exact in our theory because \( U_{00}({\mathbf{G}}) \)
depends on $\delta n$ and is reevaluated self-consistently in every
iteration. However, since the density modulations are small, we expect
that a good approximation to \( U_{00}({\mathbf{G}}) \) can be
obtained by keeping only the \( \delta n \)-independent terms given by
Equations (\ref{HF1}) and (\ref{HF4}) in the HF Hamiltonian (the term
given by Eq. (\ref{HF1}) is a constant). In that case the approximate
effective potential can be expressed as \( U_{00}({\mathbf{G}})\equiv
W_{0}({\mathbf{G}})\left\langle \Delta
_{00}({\mathbf{G}})\right\rangle \), defining the effective
interaction \( W_{0}({\mathbf{G}}) \). We display the plot of this
effective interaction in Fig. 2 for different values of the parameter
\( \Lambda \). Whereas for the \( \nu ^{*}=1/3 \) triangular lattice
we expect a minimum at about \( \left| {\mathbf{q}}\right|
l^{*}_{0}\approx 1.56 \), the minimum for CF effective potential is at
much smaller wave-vector, more so for a small \( \Lambda \). This is
why we do not limit ourselves to the triangular lattice but calculate
the ground state energies along with the \( E_{g} \)'s for three
oblique (including triangular) lattices. The results together with the
primitive reciprocal lattice vectors \( {\mathbf{b}}_{1} \), \(
{\mathbf{b}}_{2} \) are given in Table 2. Every lattice is rescaled by
an overall factor that makes the volume of the unit cell a constant
equal to \( 2\pi l^{2}_{0}/\nu \).

Table 2 {\( l=2 \) CF lattice ground state and activation energies 
for different values of \( \Lambda \) and different unit cells.}

\vspace{0.3cm}
{\centering \begin{tabular}{|c|c|c|c|c|c|}
\hline 
&
\( \Lambda  \)&
\( 0 \)&
\( l_{0}/2 \)&
\( l_{0} \)&
\( 3l_{0}/2 \)\\
\hline 
\( {\mathbf{b}}_{1}=(1,0) \)&
\( E_{HF} \)&
\multicolumn{1}{|r|}{\( -0.33 \)}&
\multicolumn{1}{|r|}{\( -0.34 \)}&
\multicolumn{1}{|r|}{\( -0.30 \)}&
\multicolumn{1}{|r|}{\( -0.24 \)}\\
\cline{2-2} \cline{3-3} \cline{4-4} \cline{5-5} \cline{6-6} 
\multicolumn{1}{|c|}{\( {\mathbf{b}}_{2}=(0.5,\sqrt{3}/2) \)}&
\( E_{g}(e^{2}/\epsilon l_{0}) \)&
\multicolumn{1}{|r|}{\( 0.08 \)}&
\multicolumn{1}{|r|}{\( 0.13 \)}&
\multicolumn{1}{|r|}{\( 0.13 \)}&
\multicolumn{1}{|r|}{\( 0.14 \)}\\
\hline 
\( {\mathbf{b}}_{1}=(1,0) \)&
\( E_{HF} \)&
\multicolumn{1}{|r|}{\( -0.36 \)}&
\multicolumn{1}{|r|}{\( -0.33 \)}&
\multicolumn{1}{|r|}{\( -0.29 \)}&
\multicolumn{1}{|r|}{\( -0.24 \)}\\
\cline{2-2} \cline{3-3} \cline{4-4} \cline{5-5} \cline{6-6} 
\multicolumn{1}{|c|}{\( {\mathbf{b}}_{2}=(0.5,\sqrt{3}) \)}&
\( E_{g}(e^{2}/\epsilon l_{0}) \)&
\multicolumn{1}{|r|}{\( 0.06 \)}&
\multicolumn{1}{|r|}{\( 0.07 \)}&
\multicolumn{1}{|r|}{\( 0.12 \)}&
\multicolumn{1}{|r|}{\( 0.12 \)}\\
\hline 
\( {\mathbf{b}}_{1}=(1,0) \)&
\( E_{HF} \)&
\multicolumn{1}{|r|}{\( -0.37 \)}&
\multicolumn{1}{|r|}{\( -0.33 \)}&
\multicolumn{1}{|r|}{\( -0.29 \)}&
\multicolumn{1}{|r|}{\( -0.24 \)}\\
\cline{2-2} \cline{3-3} \cline{4-4} \cline{5-5} \cline{6-6} 
\multicolumn{1}{|c|}{\( {\mathbf{b}}_{2}=(0.5,3\sqrt{3}/2) \)}&
\( E_{g}(e^{2}/\epsilon l_{0}) \)&
\multicolumn{1}{|r|}{\( 0.04 \)}&
\multicolumn{1}{|r|}{\( 0.05 \)}&
\multicolumn{1}{|r|}{\( 0.09 \)}&
\multicolumn{1}{|r|}{\( 0.10 \)}\\
\hline 
\end{tabular}\par}
\vspace{0.3cm}

We find that for \( \Lambda =0 \) composite fermions prefer the elongated lattices
to the triangular one. For larger values of \( \Lambda  \) the triangular lattice
is favored. The results for the excitation energy are somewhat closer to the
experimental value but still too large. 

The disagreement between the theory and the experiment is not only in the magnitude
of the activation energy but also in its dependence on the filling factor. In
our theory with \( l=2 \) the function \( E_{g}(\nu ^{*}) \) varies slowly
and is monotonic around the CF filling factor \( \nu ^{*}=1/3 \) (\( \nu =1/5 \)).
Fig. 3 gives this dependence for the triangular lattice with \( \Lambda =3l_{0}/2 \).
The experimental function (see Fig. 3 in \cite{jiang2}) has a sharp peak between
the filling factors \( \nu \sim 0.22 \) and \( \nu \sim 0.21 \) and for \( \nu <1/5 \)
it rises sharply and saturates at lower filling factors. 

Let us turn to CFs with four flux quanta to see how the results compare with experiments. 

\subsection{Crystals of Composite Fermions with four vortices attached}

The behavior of the experimental gap with $\nu$ fits in more naturally
within the CF model with \( l=4 \). When the filling factor \( \nu
<1/5 \) the lowest CF Landau level is being populated and a CF
quasiparticle lattice is assumed to be the stable
state\cite{note6}. On the other hand when $\nu>1/5$ the second CF
Landau level is being populated, and one naturally expects some
difference in the behavior of the gap in the theory. We will see that
this expectation is realized, but not in complete agreement with
experiments.

Numerical constraints
limited our HF basis to the two lowest CF Landau levels (\( n_{m}=2 \)). The initial
seed used in the HF procedure that converged to the correlated WC was
\[
\left\langle \Delta _{n_{1}n_{2}}({\mathbf{G}})\right\rangle =\left\{ \begin{array}{cc}
\nu ^{*}e^{-{\mathbf{G}}^{2}l^{*2}_{0}/4} & \textrm{if}\: n_{1},n_{2}=0\\
0 & \textrm{otherwise}.
\end{array}\right. \]
When \( \nu >1/5 \), the second CF Landau level is partially filled. Again
we assume that the CF lattice is formed so the initial seed that we use in this
case is
\[
\left\langle \Delta _{n_{1}n_{2}}({\mathbf{G}})\right\rangle =\left\{ \begin{array}{cc}
e^{-{\mathbf{G}}^{2}l^{*2}_{0}/4} & \textrm{if}\: n_{1},n_{2}=0\\
(\nu ^{*}-1)e^{-{\mathbf{G}}^{2}l^{*2}_{0}/4} & \textrm{if}\: n_{1},n_{2}=1\\
0 & \textrm{otherwise}.
\end{array}\right. \]
Our results for the activation energy are presented in Fig. 1. The value of
the parameter \( \Lambda  \) is \( 3l_{0}/2 \) (the results for \( \Lambda =l_{0} \)
are very similar) and we assume that the lattice is triangular. A magnetic field
of \( B=20 \) T was used to convert the energy units to Kelvin, in order to
compare to the work of Jiang et al\cite{jiang2}.

First we will discuss the results for \( \nu <1/5 \). The excitation
gaps that we obtain are generally comparable to the experimental
values. We also reproduce a correct \( E_{g}(\nu ) \) dependence here
(see the left half of the Fig.  1). However we do not observe
saturation towards the lower filling factors.  This may be an
indication that perhaps CFs with \( l=4 \) are not the quasiparticles
at very low fillings.

Another experimental probe supporting the crystalline nature of the insulating
state is a \( I-V \) measurement\cite{engel}. Nonlinear \( I-V \) curves
have a threshold voltage at which the differential resistance drops off that
can be interpreted as a depinning of a weakly pinned Wigner Crystal\cite{fukuyama-lee}.
As the filling factor is varied the threshold voltages increase approaching
the FQH state at \( \nu =1/5 \) both from above and below. This finding could
be a consequence of a lattice getting less rigid as the FQH state is closer\cite{fukuyama-lee,blatter}.
We have calculated the shear modulus of the CF lattice for several fractions
\( \nu <1/5 \). We first compute the ground state energies of a triangular lattice
with the primitive reciprocal lattice vectors \( {\mathbf{b}}_{1}=(1,0) \), \( {\mathbf{b}}_{2}=(0.5,\sqrt{3}/2) \)
and a deformed lattice such that it primitive reciprocal lattice vectors are
\( {\mathbf{b}}_{1}=Q_{0}(1,0) \), \( {\mathbf{b}}_{2}=Q_{0}(0.5,3\sqrt{2}/4) \)
(oblique lattice), with \( Q_{0} \) chosen so that the area of the Brillouin
zone is equal to that of the triangular lattice. Then the shear modulus \( \mu  \)
is proportional to the difference of the ground state energies. The results
are presented in Figure 4 for \( \Lambda =1.5l_{0} \). We observe that the
lattice is indeed becoming softer as $\nu\to 1/5$. This conclusion
is consistent with the experimental results\cite{engel}  interpreted
using the collective pinning theory\cite{fukuyama-lee,blatter}.

For \( \nu >1/5 \) the gaps for the triangular lattice, while being in
the same range as their experimental counterparts, do not show the
correct dependence on $\nu$ close to \( \nu =1/5 \) (see the right
half of the Fig. 1). We find that in this case the triangular lattice
is not the lowest energy solution to the HF equations.  Fig. 5 gives
the HF energies of several lattices for a fraction \( \nu ^{*}=6/5 \)
(that corresponds to \( \nu =0.206\ldots \)). The lattices that we
consider are deformations of the triangular lattice obtained from it
by changing the angle \( \theta \) between the reciprocal lattice
vectors \( {\mathbf{b}}_{1} \) and \( {\mathbf{b}}_{2} \) such that \(
\left| {\mathbf{b}}_{1}\right| =\left| {\mathbf{b}}_{2}\right| \) and
the volume of the unit cell remains a constant. Fig. 6 gives the
corresponding activation energies. As is apparent from Fig. 5, the
lattices with small angles \( \theta \) are more stable within our HF
scheme. The smallest angle for which the iterations reliably converge
is \( \theta =\pi /6 \). The dependence of \( E_{g}\) on $\nu$ is
presented for the triangular and the $\theta=\pi/6$ lattices in Fig. 1.

At this point we have run into an intrinsic limitation of the
Hamiltonian theory: Since the exact transformation between the
electronic coordinates and the CF coordinates in not known, the
Hamiltonian itself is not known exactly. This means that we should not
take the ground state energies that are predicted by our theory too
seriously. Note also that the differences in ground state energy
between the different lattice structures are very tiny so any
conclusion concerning the stability of one lattice compared to another
should be taken with a grain of salt. We still can estimate the
``shear modulus'' for this class of lattices by looking at the
difference in ground state energy between the triangular and square
lattices. This leads to an estimate of \( \mu \approx 2\times
10^{-5}\frac{e^{2}}{\epsilon l_{0}} \), an order of magnitude smaller
than for \( \nu <1/5 \). This means that the CF lattices are very soft
for $\nu$ just above 1/5, and disorder may potentially be very
important in this case. As the filling factor increases the situation
remains qualitatively similar but the differences in energy
decrease. The HF energies for the triangular and square lattices are
presented for several filling factors in Table 3. The activation gaps
for these two lattices are essentially the same.

Table 3 {Comparison of \( l=4 \) CF square and hexagonal lattice ground states for 
filling factors \( \nu > 1/5 \). Energy in units of \( \frac{e^{2}}{\epsilon l_{0}} \).}

\vspace{0.3cm}
{\centering \begin{tabular}{|c|c|c|c|c|c|}
\hline 
\( \nu  \)&
\( 0.2069 \)&
\( 0.2105 \) &
\( 0.2143 \)&
\( 0.2174 \)&
\( 0.2195 \)\\
\hline 
\hline 
\( E_{HF} \) square&
\multicolumn{1}{|r|}{\( -0.265046 \)}&
\multicolumn{1}{|r|}{\( -0.265242 \)}&
\multicolumn{1}{|r|}{\( -0.265716 \)}&
\multicolumn{1}{|r|}{\( -0.266248 \)}&
\( -0.266659 \)\\
\hline 
\( E_{HF} \) hexag.&
\multicolumn{1}{|r|}{\( -0.265040 \)}&
\multicolumn{1}{|r|}{\( -0.265232 \)}&
\multicolumn{1}{|r|}{\( -0.265708 \)}&
\( -0.266247 \)&
\( -0.266658 \)\\
\hline 
\end{tabular}\par}
\vspace{0.3cm}

\section{CONCLUSIONS, CAVEATS, AND OPEN QUESTIONS}

Two-dimensional electron gases in high magnetic fields offer the best
conditions for the realization of the Wigner Crystal, since the
magnetic field tends to localize the electrons. However, electronic
correlations play a dominant role in the LLL because the kinetic energy
is degenerate. Attempts at describing the Wigner Crystal using
uncorrelated, or weakly correlated states of electrons\cite{lam}, do capture
some of the essential physics, such as the filling factor at which
the Laughlin liquid becomes unstable to the Wigner Crystal. However,
these theories fail to capture the correct structure of the excitation
spectrum, and predict gaps that are two orders of magnitude above
experimental observations.

Since Laughlin-Jastrow correlations are the essence of the fractional
quantum Hall liquid states\cite{laugh,jain}, it is natural to
hypothesize that they are important in the Wigner Crystal state as
well. The first step in this direction was taken by Yi and
Fertig\cite{yi-fertig}, who studied the ground state energy as more
and more vortices were attached to electrons forming a Wigner
Crystal. They found that indeed the Wigner Crystals with vortices had
better energies than the uncorrelated or weakly correlated
crystals\cite{yi-fertig}.

Unfortunately, ground state energies cannot be probed directly in
experiments. It is desirable to have predictions for observable
physical properties that can distinguish between competing ground
states. Calculating physical properties in a strongly correlated state
is notoriously difficult. The Composite Fermion picture\cite{jain}
achieves the miracle of transforming a strongly correlated electronic
problem into a weakly correlated problem of CFs. In the years since
the discovery of the FQHE, much progress has been made in developing
field-theoretic schemes which have predictive
power\cite{field-bos,field-theory}. The latest in this long line of
approaches is the Hamiltonian formalism\cite{shankar-murthy}, which
has had reasonable success in computing gaps, magnetoexciton
dispersions, and finite temperature properties for the liquid
states\cite{sm-work}.

In this paper we have partially accomplished the goal of computing the
physical properties of a strongly correlated Wigner Crystal. Based on
an extension of the Hamiltonian theory to account for the nonuniform
density in the crystal state, we were able to compute gaps in the
correlated crystal. 

Our results show that qualitatively and semi-quantitatively, a Wigner
Crystal state of CFs with four flux quanta attached offers the best
description of the phenomenology of the high-field Wigner Crystal near
$\nu=1/5$. In particular, our predictions for gaps are within a factor
of 2 of the experiments in the entire regime of interest. Our
predictions show a different behavior for $\nu<1/5$ and
$\nu>1/5$. While the theory has some discrepancies with the
data\cite{jiang2} for $\nu$ just above 1/5, we believe we understand
why this might be the case: Different lattice structures have very
similar energies in this regime, and are very deformable. Consequently
disorder is expected to play a dominant role in determining the
configuration, and hence the gaps, in this region of $\nu$. Finally,
we are able to estimate the shear modulus of the crystal above and
below $1/5$, and we find them to become softer as 1/5 is
approached. This is consistent with the {\it increase} of the
threshold voltage for nonlinear transport\cite{engel}, a standard
feature of the theory of collective
pinning\cite{fukuyama-lee,blatter}.

Before we close, some caveats must be noted. An intrinsic limitation
of the Hamiltonian theory\cite{shankar-murthy} is that the Hamiltonian is known only
approximately. Thus the ground state energies are not to be taken too
seriously. This implies that this theory does not offer a trustworthy
way to find the lowest energy state. The strength of the Hamiltonian
theory lies in the fact that if the nature of the state is known, the
theory allows the computation of gaps, magnetoexcitons, and even
finite temperature properties\cite{sm-work}. With this in mind, let us
note that we have not carried out an exhaustive search in the space of
all possible states. We have confined ourselves to crystals with one
CF per unit cell. While we did explore crystals other than triangular
and square for $\nu>1/5$, we kept the two primitive reciprocal lattice
vectors equal in magnitude. It is possible that some other crystal
state that we have not explored is the actual ground state in the
clean limit. However, for the experimental observations this point is
likely to be moot, because disorder probably plays a dominant role in
this region of $\nu$.

Many open questions remain. The most important, and the most
intractable, is the influence of disorder. Disorder will cause lattice
deformations, dislocations and other defects. In a crystal of CFs,
density variations are expected to produce a corresponding variation
in the effective magnetic field. Thus, a random potential leads
indirectly to a random magnetic field. In principle, the formalism we
have developed here to deal with nonuniform density could be
generalized to incorporate disorder, but the implementation appears
difficult. In particular, it is difficult to visualize how the nonperturbative
effects of disorder (localization of almost all states, changing
$\sigma_{xy}$ from $\nu e^2/h$ to 0, etc.) would emerge in a
straightforward manner.

Another open question is the evolution of the Wigner Crystal state
with temperature, which could be explored in the clean limit along the
line of reasoning laid out in this work. In particular, it would be of
interest to obtain a prediction for the transition temperature between
the Wigner crystal and the (presumably liquid) high-temperature
state.

We hope to pursue these and other topics in future work. 

\section{Acknowledgements}
We are grateful to the NSF for partial support of this work under
grants DMR-9870681 (R.N. and H.A.F) and DMR-0071611 (R.N. and G.M.), and the
Center for Computational Sciences (R.N.) at the University of
Kentucky.

\section{APPENDIX I}

In this appendix we will construct the canonical transformation that diagonalizes
the Hamiltonian (\ref{RPA Hamiltonian}) as discussed in the main text. First
notice that putting every \( \delta n({\mathbf{G}}) \) to zero takes us back
to the uniform charge density case considered by Murthy and Shankar\cite{M-S jain book}.
They showed that the canonical transformation in that case is 
\begin{equation}
\label{0 canonical transformation }
U_{0}(\lambda )\equiv e^{i\lambda S_{0}}=e^{\left( \lambda \theta \sum _{{\mathbf{q}}}^{Q}\left( c^{\dagger }({\mathbf{q}})A({\mathbf{q}})-\textrm{h}.\textrm{c}.\right) \right) },
\end{equation}
 where \( \theta =1/2n\sqrt{\pi l} \) and \( \lambda =\arctan \mu /\mu  \)
with \( \mu ^{2}=1/l\nu -1 \). The value of the constant \( \lambda  \) is
fixed by requiring that the Hamiltonian in the FR doesn't have a term coupling
the particle and the oscillator degrees of freedom.

The Hamiltonian (\ref{RPA Hamiltonian}) is different from the one considered
in \cite{M-S jain book} by having terms proportional to \( \delta n/n \).
The same is true for the commutator of the kinetic momenta (\ref{c commutator}).
We will assume that \( \delta n/n \) is a small parameter and when diagonalizing
the Hamiltonian we will only keep terms proportional to it. Consistent with
this program a reasonable guess for the canonical transformation is 
\begin{equation}
\label{1 canonical transformation}
U(\lambda )=e^{\left( i\lambda S_{0}+\lambda \beta \sum _{{\mathbf{q}}}^{Q}\sum _{{\mathbf{G}}}^{Q}\left( c^{\dagger }({\mathbf{q}})A({\mathbf{q}}-{\mathbf{G}})\delta n({\mathbf{G}})\hat{q}_{-}\widehat{(q-G)}_{+}-\textrm{h}.\textrm{c}.\right) \right) }.
\end{equation}
 \( \beta  \) is a constant that has to be determined later by requiring that
the first order of the coupling term be zero. In (\ref{0 canonical transformation })
\( c({\mathbf{q}}) \) is the operator that corresponds to the uniform density,
while in (\ref{1 canonical transformation}) it depends on \( \delta n \).
Strictly speaking these are two different operators.

We proceed as in \cite{M-S jain book} by determining the operators \( A({\mathbf{q}},\lambda ) \)
and \( c({\mathbf{q}},\lambda ) \) in the new representation. Each of these operators
is a sum of an unperturbed part that coincides formally with \( \delta n=0 \)
result and a first order in \( \delta n/n \) part. We introduce the notation
\( A_{0}({\mathbf{q}},\lambda )+A_{1}({\mathbf{q}},\lambda ) \) for these parts
(similarly for \( c({\mathbf{q}},\lambda ) \)). Using the canonical transformation
(\ref{1 canonical transformation}) and the commutation relations (\ref{A commutator})
and (\ref{c commutator}) we derive the following first order flow equations
for the operators 
\begin{eqnarray}
\frac{dA_{1}({\mathbf{q}},\lambda )}{d\lambda } & = & -\theta c_{1}({\mathbf{q}},\lambda )\nonumber \\
 & - & \beta \sum _{{\mathbf{G}}}^{Q}c_{0}({\mathbf{q}}-{\mathbf{G}},\lambda )\delta n({\mathbf{G}})\hat{q}_{-}\widehat{(q-G)}_{+},\label{flow equation A} \\
\frac{dc_{1}({\mathbf{q}},\lambda )}{d\lambda } & = & 2eB^{*}n\theta A_{1}({\mathbf{q}},\lambda )+\theta \left( 2eB^{*}(\beta n+1)-4\pi ln\right) \nonumber \\
 & \times  & \sum _{{\mathbf{G}}}^{Q}A_{0}({\mathbf{q}}-{\mathbf{G}},\lambda )\delta n({\mathbf{G}})\hat{q}_{-}\widehat{(q-G)}_{+}.\label{flow equation c} 
\end{eqnarray}
 Substituting Eq. (\ref{flow equation c}) into Eq. (\ref{flow equation A})
a second order inhomogeneous ordinary differential equation is obtained for
\( A_{1}({\mathbf{q}},\lambda ) \). The general solution depends on two arbitrary
constants that are determined through the initial conditions \( A_{1}({\mathbf{q}},0)=c_{1}({\mathbf{q}},\lambda )=0 \).
The result of the calculation is 
\begin{eqnarray}
A_{1}({\mathbf{q}},\lambda )=-\frac{\alpha \lambda \sin \mu \lambda }{2\mu }\sum _{{\mathbf{G}}}^{Q}A({\mathbf{q}}-{\mathbf{G}})\delta n({\mathbf{G}})\hat{q}_{-}\widehat{(q-G)}_{+} &  & \nonumber \\
+\left( (\frac{\theta }{2\mu n}-\frac{\pi l\theta }{\mu eB^{*}})\sin \mu \lambda -\frac{\alpha \theta \lambda \cos \mu \lambda }{2\mu ^{2}}\right) \sum _{{\mathbf{G}}}^{Q}c({\mathbf{q}}-{\mathbf{G}})\delta n({\mathbf{G}})\hat{q}_{-}\widehat{(q-G)}_{+}, &  & \label{A1} 
\end{eqnarray}
 where a new constant \( \alpha =2\mu ^{2}(\beta /\theta +\pi l/eB^{*}-1/2n) \)
was introduced. Having found \( A_{1}({\mathbf{q}},\lambda ) \), we can integrate
\( c_{1}({\mathbf{q}},\lambda ) \) from the Eq. (\ref{flow equation c}) with
the result 
\begin{eqnarray}
c_{1}({\mathbf{q}},\lambda )=-\frac{\alpha \lambda \sin \mu \lambda }{2\mu }\sum _{{\mathbf{G}}}^{Q}c({\mathbf{q}}-{\mathbf{G}})\delta n({\mathbf{G}})\hat{q}_{-}\widehat{(q-G)}_{+} &  & \nonumber \\
+\left( (\frac{\mu }{2\theta n}-\frac{\pi l\mu }{\theta eB^{*}})\sin \mu \lambda +\frac{\alpha \lambda \cos \mu \lambda }{2\theta }\right) \sum _{{\mathbf{G}}}^{Q}A({\mathbf{q}}-{\mathbf{G}})\delta n({\mathbf{G}})\hat{q}_{-}\widehat{(q-G)}_{+}. &  & \label{c1} 
\end{eqnarray}
 The only term in the Hamiltonian (\ref{RPA Hamiltonian}) that is not expressed
through the operators \( A({\mathbf{q}}) \) and \( c({\mathbf{q}}) \) is the CP
kinetic energy \( T=\sum _{i}^{N}{\mathbf{\Pi}} _{j}^{2}/2m \). We will compute this operator
in FR by deriving first the flow equation for it. First though we have to rearrange
\( T \), by using the canonical momentum commutator, into
\begin{equation}
\label{kinetic energy}
T=\sum _{j}^{N}\frac{\Pi _{j-}\Pi _{j+}}{2m}+\sum _{j}^{N}\left( \frac{eB^{*}}{2m}-\frac{\pi l}{m}\delta n({\mathbf{r}}_{j})\right) .
\end{equation}
 The second term in (\ref{kinetic energy}) will not contribute to the flow
equation after applying the RPA approximation to it. It turns out to describe
the magnetic moment of the CP. After doing the appropriate commutators we find
that to first order in \( \delta n/n \) the kinetic energy operator \( T_{1} \)
obeys the flow equation 
\begin{eqnarray}
\frac{dT_{1}(\lambda )}{d\lambda }=\frac{eB^{*}\theta }{m}\sum _{{\mathbf{q}}}^{Q}(A^{\dagger }_{1}({\mathbf{q}},\lambda )c_{0}({\mathbf{q}},\lambda )+A^{\dagger }_{0}({\mathbf{q}},\lambda )c_{1}({\mathbf{q}},\lambda )+\textrm{h}.\textrm{c}.) &  & \nonumber \\
+\frac{eB^{*}\beta -2\pi l\theta }{m}\sum _{{\mathbf{q}}}^{Q}\sum _{{\mathbf{G}}}^{Q}\left( A^{\dagger }_{0}({\mathbf{q}},\lambda )c_{0}({\mathbf{q}}-{\mathbf{G}},\lambda )\delta n({\mathbf{G}})\hat{q}_{-}\widehat{(q-G)}_{+}+\textrm{h}.\textrm{c}.\right) . &  & \label{flow equation T} 
\end{eqnarray}
 We can integrate the kinetic energy from (\ref{flow equation T}) using the
initial condition \( T_{1}(0)=0 \). The resulting expression for the kinetic
energy in FR is substituted together with the operators \( A({\mathbf{q}},\lambda ) \)
and \( c({\mathbf{q}},\lambda ) \) in FR into the Hamiltonian (\ref{RPA Hamiltonian}).
We fix the constant \( \beta  \) by requiring that there be no coupling between
the particle and the oscillator degrees of freedom. That way we get 
\begin{equation}
\label{beta}
\beta =-\frac{\mu +(\mu ^{2}-1)\arctan \mu }{4n^{2}\mu ^{2}\sqrt{\pi l}\arctan \mu }.
\end{equation}
 The other consequences of this transformation are given in the main text.

\section{APPENDIX II}

To calculate the matrix elements \( \left\langle n_{1}\right| \widetilde{\rho }_{0}({\mathbf{q}})\left| n_{2}\right\rangle  \)
and \( \left\langle n_{1}\right| \widetilde{\rho }_{1}({\mathbf{q}},{\mathbf{G}})\left| n_{2}\right\rangle  \)
one needs to know what the corresponding matrix elements for the operators \( e^{i{\mathbf{q}}\cdot {\mathbf{r}}} \),
\( {\mathbf{q}}\times {\mathbf{\Pi}} ^{*}e^{i{\mathbf{q}}\cdot {\mathbf{r}}} \) and \( {\mathbf{q}}\times {\mathbf{\Pi}} ^{*}e^{i({\mathbf{q}}-{\mathbf{G}})\cdot {\mathbf{r}}} \)
are. We will give here only the final expressions for these matrix elements,
as the first two were derived in several papers (see for example reference\cite{sm-work})
and the third can be found using a similar approach. The matrix elements are
\begin{eqnarray}
 & \left\langle n_{1}\right| e^{i{\mathbf{q}}\cdot {\mathbf{r}}}\left| n_{2}\right\rangle  & =\sqrt{\frac{n_{2}!}{n_{1}!}}\left( \frac{i}{\sqrt{2}}(q_{x}+iq_{y})l_{0}^{*}\right) ^{n_{1}-n_{2}}\nonumber \\
 &  & \times L_{n_{2}}^{n_{1}-n_{2}}(q^{2}l_{0}^{*2}/2)e^{-q^{2}l_{0}^{*2}/4},\label{matrix element1} \\
 & l_{0}^{*2}\left\langle n_{1}\right| {\mathbf{q}}\times {\mathbf{\Pi}} ^{*}e^{i{\mathbf{q}}\cdot {\mathbf{r}}}\left| n_{2}\right\rangle  & =i\sqrt{\frac{n_{2}!}{n_{1}!}}\left( \frac{i}{\sqrt{2}}(q_{x}+iq_{y})l_{0}^{*}\right) ^{n_{1}-n_{2}}\nonumber \\
 &  & \times \left( -L_{n_{2}}^{n_{1}-n_{2}}(q^{2}l_{0}^{*2}/2)-n_{1}L_{n_{2-1}}^{n_{1}-n_{2}}(q^{2}l_{0}^{*2}/2)\right. \nonumber \\
 &  & \left. +(n_{2}+1)L^{n_{1}-n_{2}}_{n_{2}+1}(q^{2}l_{0}^{*2}/2)\right) e^{-q^{2}l_{0}^{*2}/4},\label{matrix element2} \\
 & l_{0}^{*2}\left\langle n_{1}\right| {\mathbf{q}}\times {\mathbf{\Pi}} ^{*}e^{i({\mathbf{q}}-{\mathbf{G}})\cdot {\mathbf{r}}}\left| n_{2}\right\rangle  & =i\sqrt{\frac{n_{2}!}{n_{1}!}}\left( \frac{i}{\sqrt{2}}\left( q_{x}-G_{x}+i(q_{y}-G_{y})\right) l_{0}^{*}\right) ^{n_{1}-n_{2}}\nonumber \\
 &  & \times \left( \frac{{\mathbf{q}}\cdot ({\mathbf{q}}-{\mathbf{G}})}{({\mathbf{q}}-{\mathbf{G}})^{2}}\left( -L_{n_{2}}^{n_{1}-n_{2}}\left( ({\mathbf{q}}-{\mathbf{G}})^{2}l_{0}^{*2}/2\right) \right. \right. \nonumber \\
 &  & -n_{1}L_{n_{2-1}}^{n_{1}-n_{2}}\left( ({\mathbf{q}}-{\mathbf{G}})^{2}l_{0}^{*2}/2\right) \nonumber \\
 &  & \left. +(n_{2}+1)L^{n_{1}-n_{2}}_{n_{2}+1}\left( ({\mathbf{q}}-{\mathbf{G}})^{2}l_{0}^{*2}/2\right) \right) +i\frac{{\mathbf{q}}\times {\mathbf{G}}}{({\mathbf{q}}-{\mathbf{G}})^{2}}(n_{1}-n_{2})\nonumber \\
 &  & \left. \times L_{n_{2}}^{n_{1}-n_{2}}\left( ({\mathbf{q}}-{\mathbf{G}})^{2}l_{0}^{*2}/2\right) \right) e^{-({\mathbf{q}}-{\mathbf{G}})^{2}l_{0}^{*2}/4}.\label{matrix element3} 
\end{eqnarray}

\section{APPENDIX III}

In this Appendix we will illustrate the calculation of the integrals over the
momentum \( {\mathbf{q}} \) that appear in the HF Hamiltonian Eq. (\ref{HF1})-(\ref{HF8}).
As an example we will take the integral that appears in the exchange contribution
of the two-body, first order in \( \delta n \) term. Other integrals are done
in a similar way. We choose to integrate the following term which is part of
Eq. (\ref{HF6}),
\begin{eqnarray}
U_{tbe} & = & \frac{g}{2S}\sum _{{\mathbf{q}}}V({\mathbf{q}})\left\langle 0\right| \widetilde{\rho }_{0}(-{\mathbf{q}})\left| 1\right\rangle \nonumber \\
 &  & \times \left\langle 0\right| \widetilde{\rho }_{1}({\mathbf{q}},{\mathbf{G}})\left| 0\right\rangle e^{-il_{0}^{*2}{\mathbf{q}}\times ({\mathbf{G}}+2{\mathbf{G}}_{1})/2}.\label{integral0} 
\end{eqnarray}
Using the formulas given in the Appendix II for the density operator matrix
elements we find

\begin{eqnarray}
U_{tbe} & = & \frac{2\pi e^{2}g}{2\epsilon }\int _{0}^{\infty }\frac{dq}{(2\pi )^{2}}\int _{0}^{2\pi }d\theta e^{-q^{2}\Lambda ^{2}}\nonumber \\
 &  & \times \left( il_{0}^{*}qe^{-i\theta }\frac{q^{2}l_{0}^{*2}}{2\sqrt{2}}e^{-q^{2}l_{0}^{*2}/4}\right) \nonumber \\
 &  & \times \left( c^{2}\frac{{\mathbf{q}}\cdot {\mathbf{G}}}{G^{2}}-\frac{c}{2(c+1)}(q^{2}l_{0}^{*2}+{\mathbf{q}}\cdot {\mathbf{G}}l_{0}^{*2})\right) \nonumber \\
 &  & \times e^{-({\mathbf{q}}-{\mathbf{G}})^{2}l_{0}^{*2}/4}e^{-il_{0}^{*2}{\mathbf{q}}\times ({\mathbf{G}}+2{\mathbf{G}}_{1})/2}.\label{integral1} 
\end{eqnarray}
Taking into account that \( {\mathbf{q}}\cdot {\mathbf{G}}=q(G_{-}e^{i\theta }+G_{+}e^{-i\theta })/2 \)
and \( {\mathbf{q}}\times {\mathbf{G}}=q(G_{+}e^{-i\theta }-G_{-}e^{i\theta })/2i \),
where \( G_{+}=G_{x}+iG_{y} \) and \( G_{-}=G_{x}-iG_{y} \), we get 
\begin{eqnarray}
U_{tbe} & = & \frac{e^{2}g}{2\epsilon l_{0}^{*}}\int _{0}^{\infty }dx\int _{0}^{2\pi }\frac{d\theta }{2\pi }e^{-x^{2}(\Lambda ^{2}/l_{0}^{*2}+1/2)}\nonumber \\
 &  & \times \left( ie^{-i\theta }\frac{x^{3}}{2\sqrt{2}}\right) \left( c^{2}\frac{x(G_{-}e^{i\theta }+G_{+}e^{-i\theta })}{2G^{2}l_{0}^{*}}\right. \nonumber \\
 &  & \left. -\frac{c}{2(c+1)}(x^{2}+x(G_{-}e^{i\theta }+G_{+}e^{-i\theta })l_{0}^{*})\right) \nonumber \\
 &  & \times e^{\left( (G_{-}+G_{1-})e^{i\theta }-G_{1+}e^{-i\theta }\right) l_{0}^{*}x/2}e^{-l_{0}^{*2}G^{2}/4},\label{integral2} 
\end{eqnarray}
where \( x=ql_{0}^{*} \). First we will integrate with respect to the variable
\( x \). We notice that it is possible to extend the interval of the integration
over the whole real axis. The integrand in Eq. (\ref{integral2}) is such that
the odd/even powers of \( x \) are multiplied by the \( \exp (i\theta n) \)
with \( n \) odd/even. Then reversing the sign of \( x \) and making a transformation
\( \theta '=\theta +\pi  \) doesn't change the integrand while shifting the
integration with respect to \( x \) interval to \( (-\infty ,0) \). The integral
is then

\begin{eqnarray}
U_{tbe} & = & \frac{ie^{2}g}{4\epsilon l_{0}^{*}}\int _{0}^{2\pi }\frac{d\theta }{2\pi }\left( I_{4}(\alpha ,\beta )\frac{(G_{-}+G_{+}e^{-2i\theta })}{4\sqrt{2}}\right. \nonumber \\
 &  & \times (\frac{c^{2}}{G^{2}l_{0}^{*}}-\frac{cl_{0}^{*}}{c+1})\left. -I_{5}(\alpha ,\beta )\frac{ce^{-i\theta }}{4\sqrt{2}(c+1)}\right) e^{-l_{0}^{*2}G^{2}/4}.\label{integral3} 
\end{eqnarray}
We introduced a notation for the Gaussian integral \( I_{n}(\alpha ,\beta )=\int ^{\infty }_{-\infty }dx\exp (-\alpha x^{2}+2\beta x)x^{n} \)
with \( \alpha =\Lambda ^{2}/l_{0}^{*2}+1/2 \) and \( \beta =\left( (G_{-}+G_{1-})e^{i\theta }-G_{1+}e^{-i\theta }\right) l_{0}^{*}/4 \).
An important observation is that after Eq. (\ref{integral3}) is expanded the
result is the sum of the integrals of the form (note the even powers of \(  e^{i\theta }  \)
that appear)
\begin{eqnarray}
\int ^{2\pi }_{0}e^{(ae^{i\theta }+be^{-i\theta })^{2}}e^{i2n\theta }\frac{d\theta }{2\pi } & = & e^{2ab}(\frac{b}{a})^{n}I_{\left| n\right| }(2ab),\nonumber \\
\int ^{2\pi }_{0}e^{(ae^{i\theta })^{2}}e^{i2n\theta }\frac{d\theta }{2\pi } & = & \left\{ \begin{array}{ll}
0 & \textrm{if}\: n>0\\
\frac{a^{\left| n\right| }}{\left| n\right| !} & \textrm{otherwise},
\end{array}\right. \nonumber \\
\int ^{2\pi }_{0}e^{(be^{-i\theta })^{2}}e^{i2n\theta }\frac{d\theta }{2\pi } & = & \left\{ \begin{array}{ll}
0 & \textrm{if}\: n<0\\
\frac{b^{n}}{n!} & \textrm{otherwise}.
\end{array}\right. \label{integral4} 
\end{eqnarray}
Here \( a=(G_{-}+G_{1-}) \) and \( b=-G_{1+} \). The second and third lines
in Eq.(\ref{integral4}) are given because they are used to calculate other
integrals. They hold if either one of \( a \) or \( b \) are zero. The final
answer is then a series of the modified Bessel functions multiplied by appropriate
constants.

\newpage

\vspace{0.3cm}
{\par\centering \resizebox*{10cm}{7cm}{\rotatebox{-90}{\includegraphics{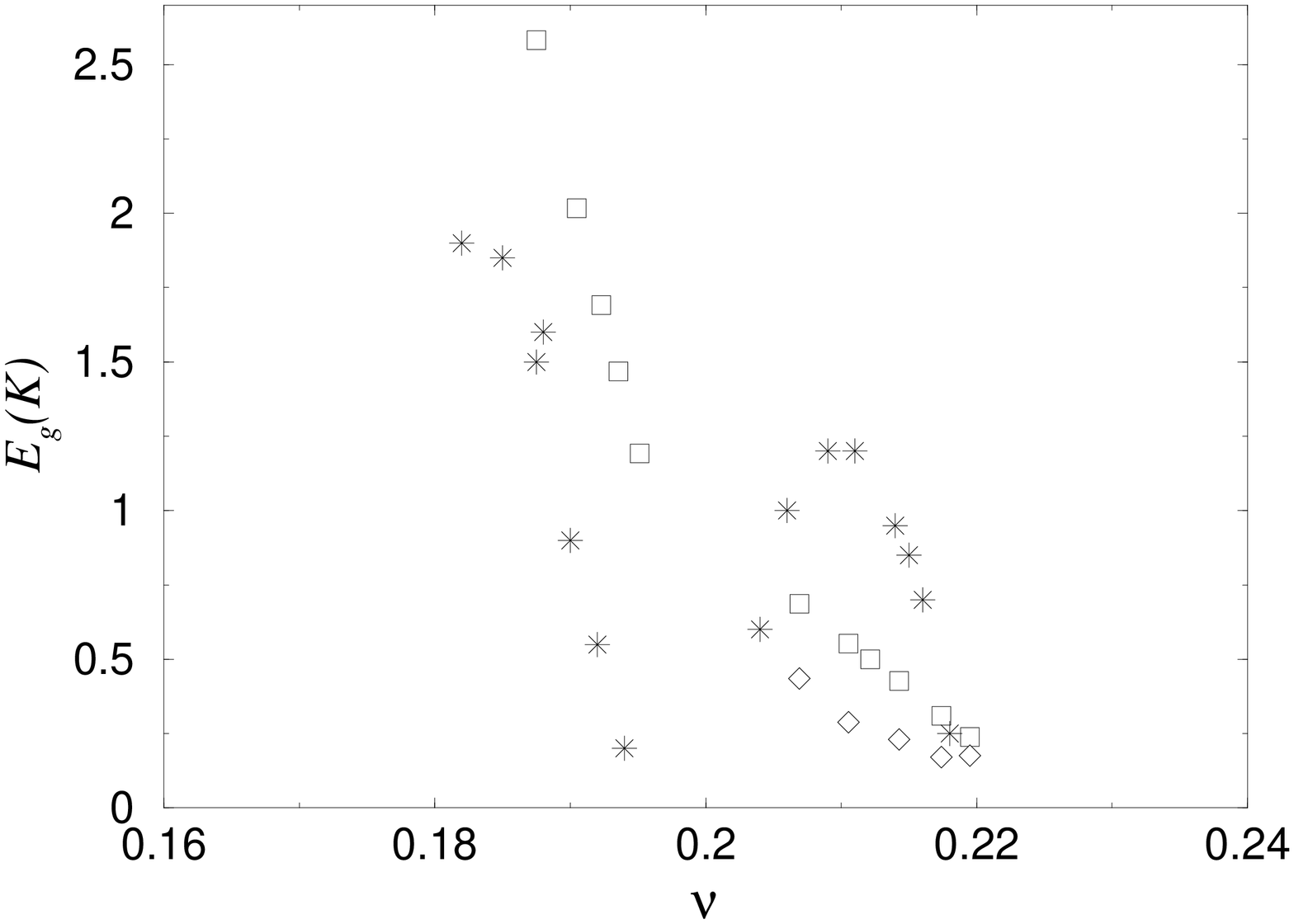}}} \par}
\vspace{0.3cm}

Figure 1.

Caption: The activation gap dependence on the filling factor around \( \nu \approx 1/5 \). Squares are our CF theory with four vortices attached 
for the hexagonal lattice. 
Diamonds represent our CF theory with four vortices attached
for the oblique lattice (see the text).
Stars are experimental results read off Fig. 3 of reference\cite{jiang2}.

\vspace{0.3cm}
{\par\centering \resizebox*{10cm}{7cm}{\rotatebox{-90}{\includegraphics{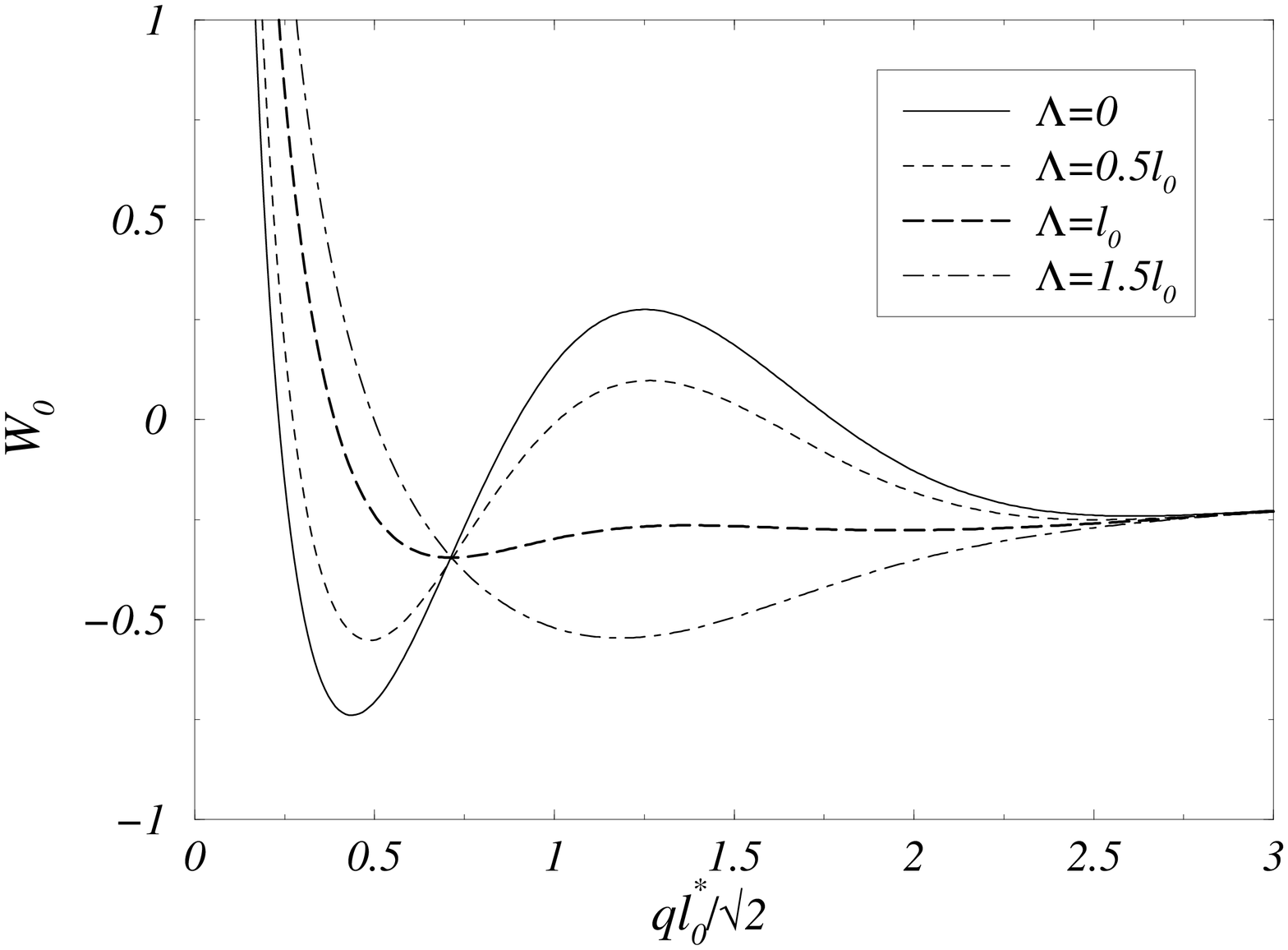}}} \par}
\vspace{0.3cm}

Figure 2.

Caption: Effective HF potential for different values of parameter \( \Lambda  \).
The filling fraction specific factor \( (1-l\nu )^{2} \) was omitted from the
zeroth-order expression of the effective potential when generating these curves. 

\vspace{0.3cm}
{\par\centering \resizebox*{10cm}{7cm}{\rotatebox{-90}{\includegraphics{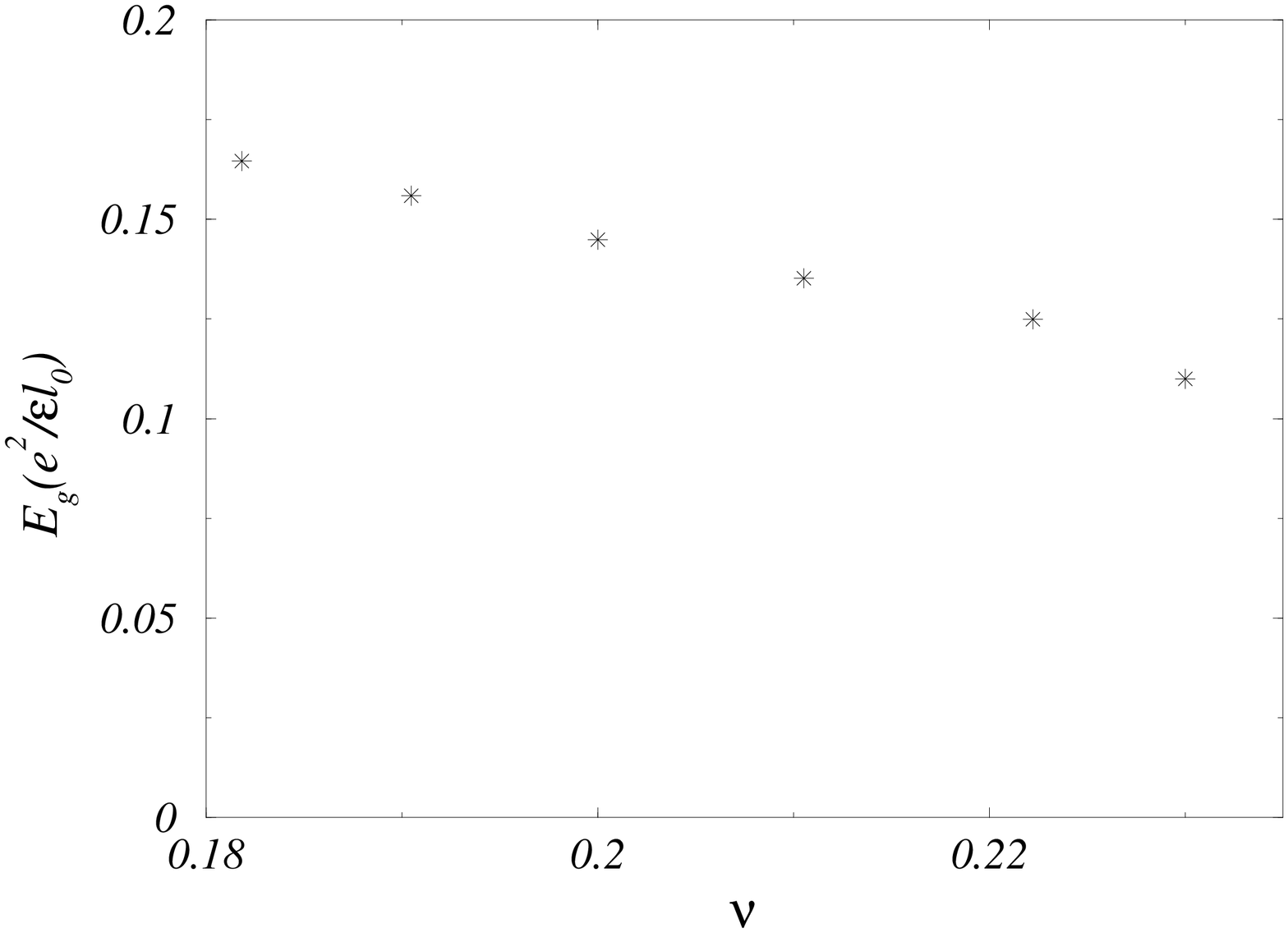}}} \par}
\vspace{0.3cm}

Figure 3.

Caption: The activation gap dependence on the filling factor around \( \nu \approx 1/5 \)
(\( l=2 \)).

\vspace{0.3cm}
{\par\centering \resizebox*{10cm}{7cm}{\rotatebox{-90}{\includegraphics{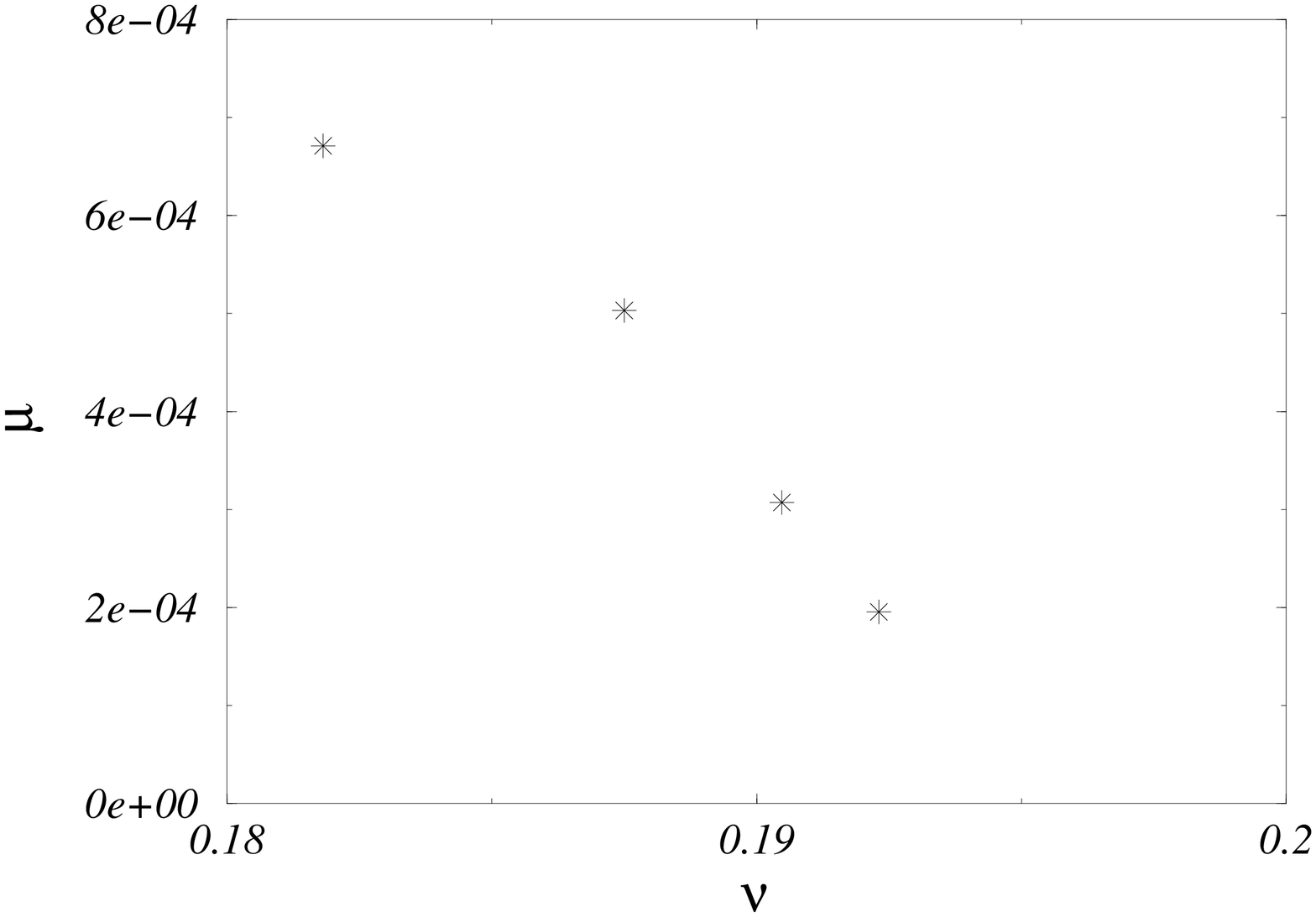}}} \par}
\vspace{0.3cm}

Figure 4.

Caption: The shear modulus \( \mu  \) for the triangular CF lattices as a function
of filling factor (\( \nu <1/5 \)).

\vspace{0.3cm}
{\par\centering \resizebox*{10cm}{7cm}{\rotatebox{-90}{\includegraphics{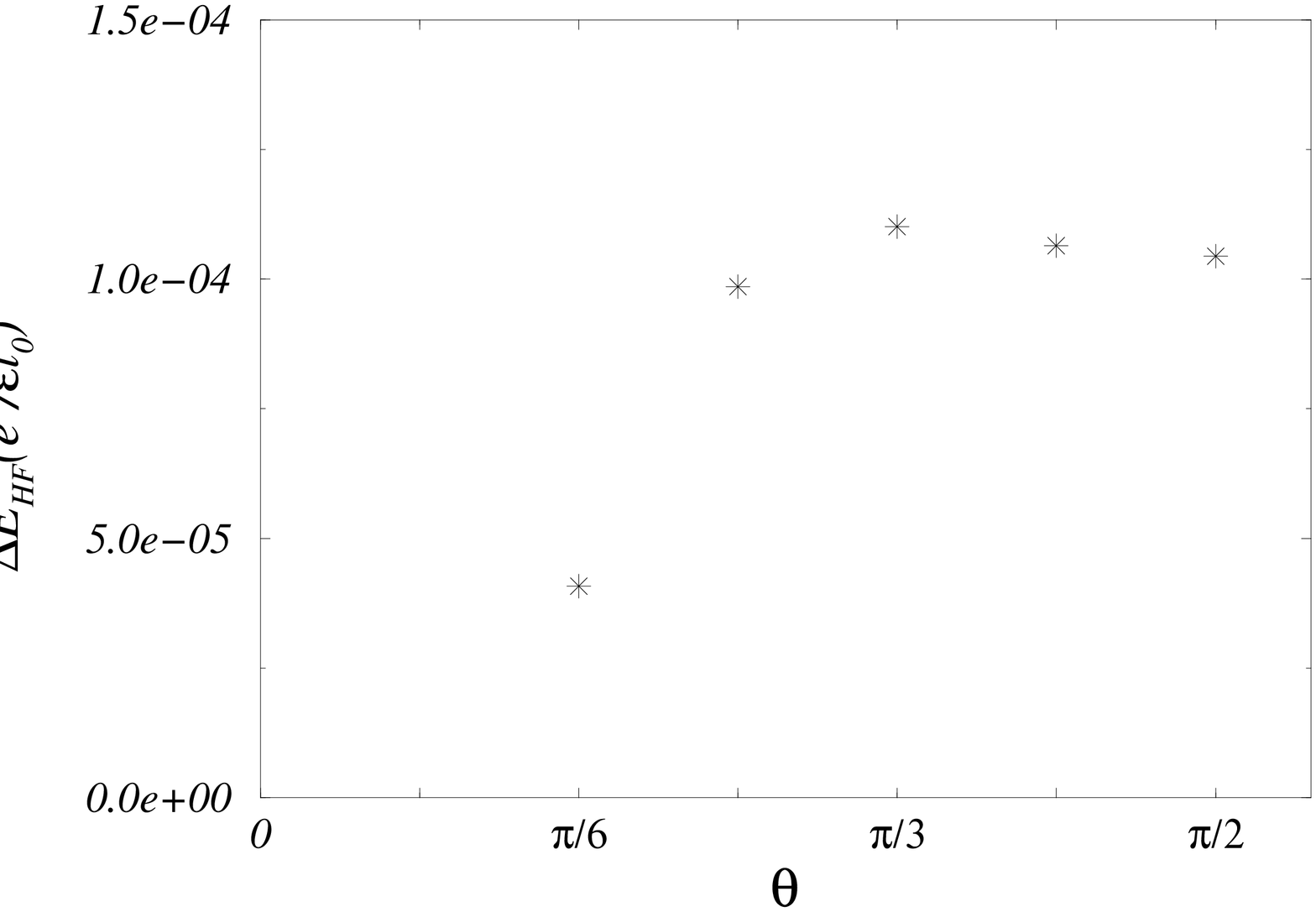}}} \par}
\vspace{0.3cm}

Figure 5.

Caption: The HF energies for CF lattices differing by an angle \( \theta  \)
between the reciprocal lattice vectors for the filling factor \( \nu \approx 0.206 \).
The zero on the vertical axis corresponds to the energy \( -0.26515\frac{e^{2}}{\epsilon l_{0}} \).

\vspace{0.3cm}
{\par\centering \resizebox*{10cm}{7cm}{\rotatebox{-90}{\includegraphics{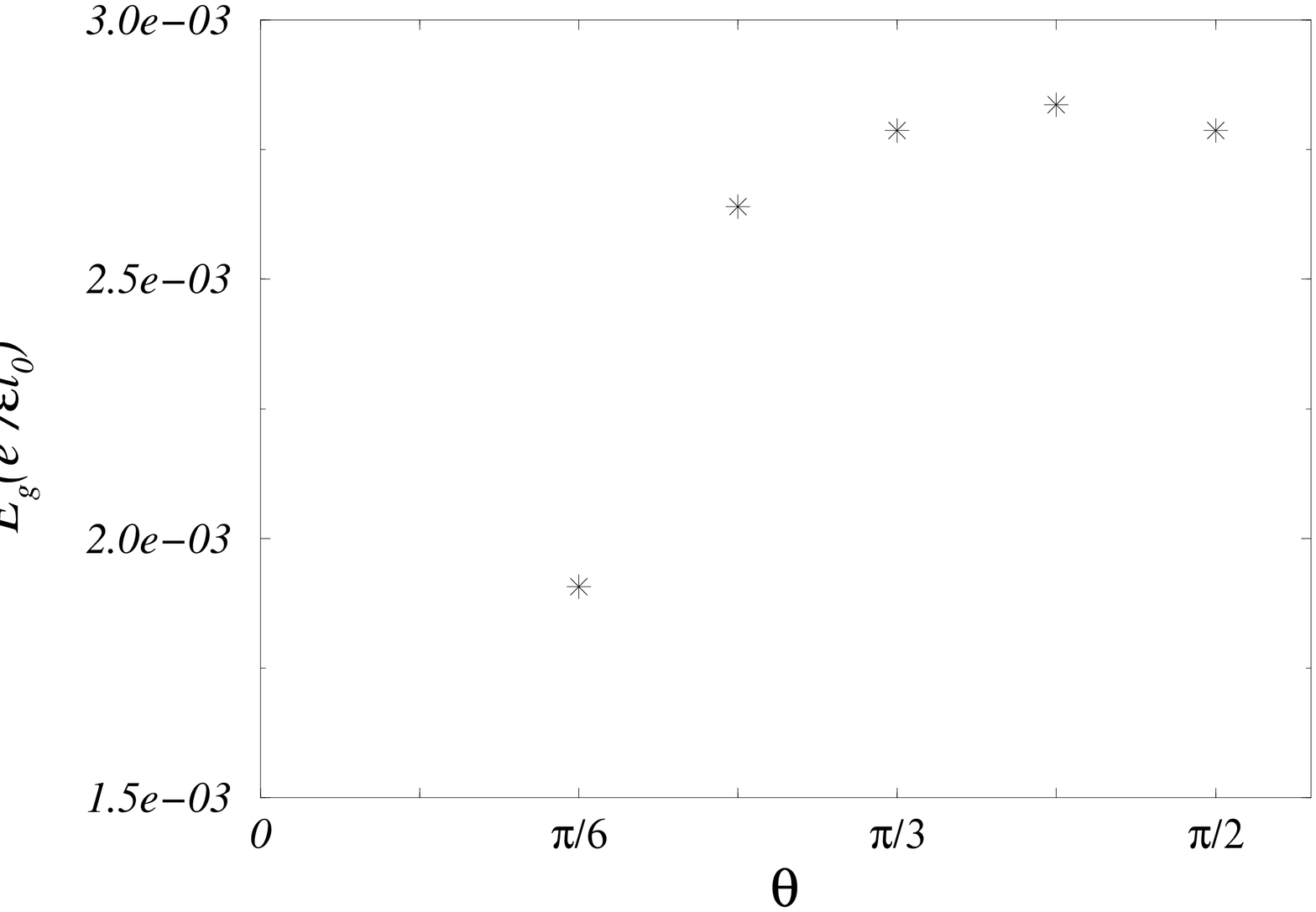}}} \par}
\vspace{0.3cm}

Figure 6.

Caption: The activation energies for CF lattices differing by an angle \( \theta  \)
between the reciprocal lattice vectors for the filling factor \( \nu \approx 0.206 \).


\begin{thebibliography}{10}
\bibitem{note1}Note that quantities of the form \( V_{\pm } \) mean \( V_{x}\pm iV_{y} \).
\bibitem{note3}Equation (\ref{density0}) differs from its uniform density counterpart\cite{M-S jain book}
because the canonical momentum operator that enters the definition of \( c({\mathbf{q}}) \)
here depends on the density modulations through the CS gauge field.
\bibitem{note4}The operator \( c({\mathbf{q}}) \) depends implicitly on \( \delta n \), because
\( {\mathbf{\Pi}} _{j} \) contains the vector potential \( {\mathbf{A}}^{*}({\mathbf{r}}_{j}) \),
which in its turn is related to the average density through \( \nabla \times {\mathbf{A}}^{*}({\mathbf{r}}_{j})=B-2\pi ln/e-2\pi l\delta n({\mathbf{r}}_{j})/e \).
\bibitem{note5}When \( M\neq 1 \) there are overlapping bands so that the bands that overlap
with \( Mq \)'th and \( Mq+1 \)'st bands should be included in determining
the excitation gap.
\bibitem{note6}We looked into the possibility that the quasihole
lattice may be formed but found that its ground state energy is higher
in the relevant range of filling factors.
\end{thebibliography}
\end{document}